\documentclass[twocolumn,prb]{revtex4}
\usepackage{amsmath,amsthm}
\usepackage{epsfig}
\usepackage{graphics}
\usepackage{graphicx}
\usepackage{dcolumn}
\usepackage{bm}
\usepackage{multirow}

\begin{document}
\title{Ferromagnetism and infrared electrodynamics of Ga$_{1-x}$Mn$_{x}$As}

\author{B. C. Chapler,$^{1,\ast}$ S. Mack,$^{2,3}$  R. C. Myers,$^4$ A. Frenzel,$^1$ B. C. Pursley,$^1$ K. S. Burch,$^5$ A. M. Dattelbaum,$^6$ N. Samarth,$^7$ D. D. Awschalom,$^2$ and D. N. Basov$^1$ }

\affiliation
{$^{1}$Physics Department, University of California-San Diego, La Jolla, California 92093, USA \\
$^{2}$Center for Spintronics and Quantum Computation, University of California-Santa Barbara, California 93106, USA (address where work of SM was completed)\\
$^{3}$ Naval Research Laboratory, Washington, DC 20375, USA (Current address for SM)\\
$^{4}$Department of Materials Science and Engineering, Ohio State University, Columbus, Ohio 43210, USA \\
$^{5}$Department of Physics \& Institute for Optical Sciences, University of Toronto, Toronto, Ontario, Canada M5S 1A7 \\
$^{6}$Los Alamos National Laboratory, Los Alamos, New Mexico 87545, USA \\
$^{7}$Department of Physics, The Pennsylvania State University, University Park, Pennsylvania 16802, USA 
}

\date{\today}

\begin{abstract} 

We report on the magnetic and the electronic properties of the prototype dilute magnetic semiconductor Ga$_{1-x}$Mn$_x$As using infrared (IR) spectroscopy. Trends in the ferromagnetic transition temperature $T_C$ with respect to the IR spectral weight are examined using a sum-rule analysis of IR conductivity spectra. We find non-monotonic behavior of trends in $T_C$ with the spectral weight to effective Mn ratio, which suggest a strong double-exchange component to the FM mechanism, and highlights the important role of impurity states and localization at the Fermi level. Spectroscopic features of the IR conductivity are tracked as they evolve with temperature, doping, annealing, As-antisite compensation, and are found only to be consistent with an Mn-induced IB scenario. Furthermore, our detailed exploration of these spectral features demonstrates that seemingly conflicting trends reported in the literature regarding a broad mid-IR resonance with respect to carrier density in Ga$_{1-x}$Mn$_x$As are in fact not contradictory. Our study thus provides a consistent experimental picture of the magnetic and electronic properties of Ga$_{1-x}$Mn$_x$As.

\end{abstract}


\maketitle

\section{Introduction}

Much of the interest in the ferromagnetic semiconductor Ga$_{1-x}$Mn$_x$As stems from the rich physics of metallicity $and$ ferromagnetism induced in a semiconducting host through doping with magnetic impurities. There is a general consensus that itinerant holes, introduced by the Mn doping, mediate the ferromagnetic interaction between the magnetic moments of the Mn ions~\cite{Jungwirth2006b, Burch2008, Sato2010}. Thus all proposed ferromagnetic (FM) mechanisms of Ga$_{1-x}$Mn$_x$As are intimately tied to the dynamics of the charge carriers. This interplay between the electronic structure and magnetism necessitates studies that can form a comprehensive description of both the observed electronic $and$ magnetic properties in a consistent manner. This need is amplified by a long standing controversy regarding the location of Fermi level ($E_F$) in the band structure of FM Ga$_{1-x}$Mn$_x$As (See reviews by  Dietl~\cite{Dietl2010}, and or Samarth~\cite{Samarth2012a}, for instance). This controversy centers on the character of the states at $E_F$, as we describe below.

In one scenario, Zener's double exchange, of which the electronic structure is diagramed in the top panel of Fig.~\ref{mega}a, $E_F$ resides in the partially occupied majority band of Mn impurity $d$ states. In this picture, electronic conduction via hopping within the impurity band (IB) mediates the magnetic exchange, with an energy gain controlled by the magnitude of the hopping matrix element $t$ for ferromagnetically coupled Mn impurities~\cite{Anderson1963, Sheu2007, Sato2010}. Because this interaction relies on itinerant carriers, there is no energy gain for ferromagnetism if the IB is completely filled or completely empty. 

In the alternative Zener's $p$-$d$ exchange scenario (middle panel of Fig.~\ref{mega}a), $E_F$ resides in the exchange split valence $p$ band of the host. In this picture hybridization of the impurity $d$ wavefunctions with the $p$ wavefunctions of the neighboring $p$ elements produce level repulsion of the like spins states. Thus the majority spin $p$ band is shifted to higher energies, and the minority band is shifted to lower energies. This scenario results in a relatively weak, yet very long ranged interaction that is due to the extended nature of the $p$ states of the host valence band (VB)~\cite{Dietl2000, Jungwirth2006b}.

Double exchange and $p$-$d$ exchange are not necessarily mutually exclusive scenarios for the true nature of the interatomic exchange interaction. Rather, double exchange and $p$-$d$ exchange may both contribute to the resultant FM ground state of Ga$_{1-x}$Mn$_x$As. In terms of the magnetic interactions, the IB double exchange and the VB $p$-$d$ exchange scenarios are merely the strong-coupling narrow-band limit, and the weak-coupling extended state limit of each other, respectively. Other magnetic exchange interactions may also play a secondary role as well, but these are beyond the scope of this paper (see reviews of Ref.~\cite{Jungwirth2006b} and Ref.~\cite{Sato2010} for a complete profile of magnetic interactions). The crucial point is that the nature of the interaction depends heavily on details of the electronic structure and degree of localization of the mediating holes. 

In this work, we experimentally address both the magnetic and the electronic properties of the prototype dilute magnetic semiconductor Ga$_{1-x}$Mn$_x$As using infrared (IR) spectroscopy. We first examine the relationship between the infrared spectral weight, which is proportional to the carrier density (Eq.~\ref{drude}), and the ferromagnetic transition temperature $T_C$. This relationship is determined through a sum-rule analysis of our data and additional IR data available in the literature (Refs.~\cite{Singley2002, Burch2006, Jungwirth2010}) (Sec.~\ref{tc}). Our analysis shows the position and degree of localization at $E_F$ plays a key role in controlling $T_C$ (Fig.~\ref{mega}d and discussion in Sec.~\ref{tc}). In Sec.~\ref{intragap}, we perform a detailed examination of the spectral features observed in the IR data of our Ga$_{1-x}$Mn$_x$As films, and show that these features are also consistent only with a Mn-induced IB scenario, and that $E_F$ resides in this region. In this latter section, we discuss a connection between the previous experimental findings of Refs.~\cite{Burch2006} and ~\cite{Jungwirth2010}, which reported seemingly conflicting trends in the peak frequency of a broad mid-IR resonance $\omega_0$ with respect to carrier density. By examining trends in $\omega_0$ with respect to temperature, As:Ga growth ratio, Mn dopant concentration, in both as-grown and annealed samples, our data show that the results of these earlier experiments are in fact not contradictory at all. Moreover, the $\omega_0$ data is consistent with excitations to Mn-induced impurity states. 

The samples interrogated in our experiments are grown using a ``non-rotated'' molecular beam epitaxy (MBE) technique. The non-rotated technique aims to reduce or eliminate As antisite (As$_\mathrm{{Ga}}$) compensation through spatial control of the As-flux during growth (Sec.~\ref{wtf}). Incorporation of compensating defects is the primary limiting factor in minimizing disorder and maximizing $p$ in Ga$_{1-x}$Mn$_x$As. The most prevalent compensating defects are Mn-interstitials (Mn$_i$), and As-antisites (As$_\mathrm{{Ga}}$), both of which are double donors~\cite{Erwin2002, Missous1994}. These defects are an unintentional result of the low growth temperatures necessary to achieve Mn concentrations sufficient for ferromagnetism and to suppress the formation of secondary phases (e.g. MnAs)~\cite{Ohno1996}. 

Post-growth annealing at relatively low temperatures has been shown to reduce Mn$_i$ because of their high diffusivity, therefore increasing the hole density $p$~\cite{Edmonds2004, Ku2003}. The As$_\mathrm{{Ga}}$ defects cannot be removed via annealing at temperatures below 500 $^{\circ}{\rm C}$. Such high annealing temperatures are impractical, however, because at these temperatures Mn also precipitates out to form MnAs nanoparticles~\cite{Boeck1996}. The necessity for high temperature annealing to reduce As$\mathrm{Ga}$ has been circumvented by a ``non-rotated''  growth technique with spatial control of the As-flux (Sec.~\ref{wtf}), which results in a film with a corresponding As$_\mathrm{{Ga}}$ gradient~\cite{Myers2006}. Investigating films grown by this technique, the authors of Ref.~\cite{Myers2006} have shown that this precise control of the As-flux leads to the systematic reduction (or elimination) of As$_\mathrm{{Ga}}$ along the gradient.

The gradient composition of these samples results in a location along the film where As$_\mathrm{{Ga}}$ defects have been minimized or eliminated (see Fig.~\ref{rawT}). Thus by experimentally probing this optimized location in a film, we obtain data representative of a relatively ``clean'' sample. Data from the optimized location in our films are reported, for instance, in our analysis of ferromagnetism in Sec.~\ref{tc}. Samples grown using the non-rotated technique can provide additional pertinent information through investigation of systematic changes to the IR conductivity spectra ($\sigma_1(\omega)$) of Ga$_{1-x}$Mn$_x$As films along the As:Ga gradient. Tuning the material properties in this manner, and tracking the evolution of the spectroscopic features across the As$_\mathrm{{Ga}}$ gradient, yields a detailed picture of the electrodynamics of Ga$_{1-x}$Mn$_x$As. Therefore our infrared study, combining an infrared spectral weight analysis of ferromagnetism and the systematic study of the infrared spectral features, serves as a unique endeavor to use a single experimental technique to address both the electronic and magnetic properties of this prototype FM semiconductor.

To further establish that the electrodynamics of FM Ga$_{1-x}$Mn$_x$As is distinct from genuine metallic behavior due to extended states in the host VB, we also examine an $x$=0.009 Ga$_{1-x}$Be$_x$As sample grown using the same non-rotated technique. This is an ideal system to compare and contrast with the electrodynamics of Ga$_{1-x}$Mn$_x$As, as substitutional Be also forms a single acceptor state in GaAs, however, without the accompanying magnetic moment. The Ga$_{0.991}$Be$_{0.009}$As sample displays conventional metallicity with spectroscopic features consistent with transport in the host VB, in contrast to that of Ga$_{1-x}$Mn$_x$As.

The paper is organized as follows. First, we present our analysis of the relationship between the IR spectral weight and $T_C$ in Sec.~\ref{tc}. Following that, we provide details of our samples and experimental methods in Sec.~\ref{wtf}. Sec.~\ref{intragap} covers the IR conductivity spectra of our $p$-doped GaAs films, Ga$_{1-x}$Mn$_x$As and Ga$_{1-x}$Be$_x$As. This latter sections examines the dependence of the spectra on the As:Ga growth ratio in both Ga$_{1-x}$Mn$_x$As and Ga$_{1-x}$Be$_x$As, and on the doping dependence in Ga$_{1-x}$Mn$_x$As. Discussion of the trends of $\omega_0$ is also found in Sec.~\ref{intragap}. Finally, concluding statements are found in Sec.~\ref{conclusion}.

\section{An infrared perspective on trends in $T_C$}
\label{tc}

In Fig.~\ref{mega}b we plot $T_C$ as a function of dopant concentration $x$ of all the Ga$_{1-x}$Mn$_x$As samples in our analysis. The samples include as-grown films investigated by Singley $et$ $al.$~\cite{Singley2002}, as-grown and annealed films of Burch $et$ $al.$~\cite{Burch2006}, annealed films of Jungwirth $et$ $al.$~\cite{Jungwirth2010}, and the non-rotated as-grown and annealed samples of this work (see Table~\ref{thickness}). For the later samples presented in Fig.~\ref{mega}, only the optimized location along the As:Ga gradient is reported in the figure (method for determination of this location is found Sec.~\ref{sectionA}). In these works, the total Mn concentration $x$ was determined by electron microprobe analysis (Refs.~\cite{Singley2002, Burch2006}), measuring the ratio of beam equivalent pressures of Mn and Ga sources before each growth (and cross checked on several samples using secondary ion mass spectroscopy) (Ref.~\cite{Jungwirth2010}), and growth rate calibrations of MnAs and GaAs reflection high-energy electron diffraction (RHEED) oscillations (this work). The totality of the data in Fig.~\ref{mega}b show no clear systematic trend across the doping profile, and non-reproducible $T_C$ for samples of the same nominal dopant concentration. These observations highlight the difficult nature of predicting $T_C$ based on Mn-concentration alone in this defect prone material. 

The relationship between carrier density and $T_C$, as opposed to $x$ and $T_C$, can provide a more insightful picture into the nature of magnetism in this carrier mediated ferromagnet. IR experiments serve as a contactless method sensitive to the carrier density through the sum-rule given by

\begin{equation}
{N_{\mathrm{eff}}=\frac{p}{m_{\mathrm{opt}}}=\frac{2}{\pi e^2}\int_{0}^{\omega_c}\sigma_{1}(\omega)d\omega}.
\label{drude}
\end{equation}

\noindent The ``spectral weight'' $N_{\mathrm{eff}}$ is proportional to the effective number of charges contributing to electromagnetic absorption at frequencies below $\omega_c$. Therefore Eq.~\ref{drude} establishes the relationship between the spectral weight $N_{\mathrm{eff}}$ and the charge density $p$. When Eq.~\ref{drude} is applied to a feature originating from an \emph{inter}band process, $m_{\mathrm{opt}}$ is the reduced mass of the bands involved~\cite{Dressel}. When  Eq.~\ref{drude} is applied to a spectral feature arising from an \emph{intra}band process, $m_{\mathrm{opt}}$ is the effective mass of the relevant band. If Eq.~\ref{drude} is integrated over multiple features arising from distinct processes, $m_{\mathrm{opt}}$ will be a ``mixture'' of the effective masses of each of the excitation processes contributing to the spectrum below $\omega_c$. 

In Fig.~\ref{mega}c we plot $T_C$ as a function of $N_{\mathrm{eff}}$ in the Ga$_{1-x}$Mn$_x$As samples. We use $\omega_c$=6450 cm$^{-1}$ as the integration cut-off in order to rule out any significant contribution from excitations into the GaAs conduction band. Moreover, this cut-off is well established in these materials and facilitates direct comparison with other studies in the literature\cite{Sinova2002, Burch2006, Chapler2011}. As can be seen in the figure, while the data do not perfectly collapse onto a singlular trend line, there is an overall trend indicating that increasing $N_{\mathrm{eff}}$ increases $T_C$. The trend in Fig.~\ref{mega}b is indicated by a power law fit (black line) of all the data points, which finds an exponent of 0.67.

Remarkably, the overall trend in Fig.~\ref{mega}c is evident despite the fact that there was no clear trend in $T_C$ versus $x$ (Fig.~\ref{mega}b), and despite the fact that the samples in Fig.~\ref{mega} span a wide variety of Mn doping levels and growth procedures, including different thicknesses, rotated and non-rotated growths, and as-grown and annealed samples (Table~\ref{thickness}). The data of Fig.~\ref{mega}b seemingly confirms earlier results of Ref.~\cite{Ku2003}, which also found an empirical relationship between $T_C$ and $p$ in Ga$_{1-x}$Mn$_x$As, apparently independent of other physical parameters (though $T_C$ scaled as $\sim p^{1/3}$ rather than $\sim N_{\mathrm{eff}}^{0.67}$ as found here). These earlier results, however, relied on a deconvolution of coupled optical phonon-plasmon modes in Raman spectra. Our result is achieved in a more straightforward manner, as a simple integration of the intragap spectra weight from our IR conductivity. The overall trend in Fig.~\ref{mega}c suggests a strong relationship between $T_C$ and $p$, which is encouraging given that ferromagnetism in Ga$_{1-x}$Mn$_x$As is widely agreed upon to be carrier mediated. Another encouraging aspect of Fig.~\ref{mega}c is that up to the largest measured spectral weight ($N_\mathrm{{eff}}>10^{21}$ cm$^{-3}/m_e$), there are no signs that the trend of increasing $T_C$ is saturating.

One obvious detail missing in any analysis of $T_C$ as a function of $p$ only, such as that described above for the data of Fig.~\ref{mega}c, is the concentration of Mn moments, which of course are necessary for ferromagnetism. Thus we are interested in estimating the effective Mn concentration $x_\mathrm{{eff}}$ of our samples. Out of the total Mn concentration $x$, the Mn will reside either in substitutional positions or interstitial positions, therefore $x=x_{sub}+x_i$, where $x_{sub}$ is the concentration of Mn$_\mathrm{{Ga}}$ and $x_i$ that of Mn$_i$. Both theory and experiments suggest that Mn$_i$--Mn$_\mathrm{{Ga}}$ pairs couple antiferromagnetically~\cite{Blinowski2003, Masek2004, Edmonds2004, Bouzerar2005, Takeda2008, Zhu2007}, therefore the effective Mn concentration contributing to ferromagnetic order is then $x_\mathrm{{eff}}=x_{sub}-x_i$. Mn$_i$ are also considered to be double donors~\cite{Erwin2002, Missous1994}, compensating the intentional hole doping of Mn$_\mathrm{{Ga}}$. Through this latter fact, and by invoking the relationship between the carrier density and the spectral weight established in Eq.~\ref{drude}, we can write that $\frac{a^3}{4}m_\mathrm{{opt}}N_{\mathrm{eff}}= x_{sub}-2x_i$, where $a$ is the GaAs lattice constant (for simplicity we have assumed the concentration of As$_\mathrm{{Ga}}$ is zero). Putting these relations together, we arrive at our estimate of effective Mn concentration $x_\mathrm{{eff}}=\frac{1}{3}x+(\frac{2}{3})\frac{a^3}{4}m_\mathrm{{opt}}N_{\mathrm{eff}}$.

For $m_\mathrm{{opt}}$ in our equation of $x_\mathrm{{eff}}$, we choose $m_\mathrm{{opt}}$ to be in the range of the effective VB mass, placing it from 0.3--0.4 $m_e$. This choice was guided by IR experiments on (Ga,Mn)As-based electric field effect devices~\cite{Chapler2012}. These latter experiments found, noting that $m_\mathrm{{opt}}$ is dominated by the VB mass the integration limit in the sum-rule (Eq.~\ref{drude}) is on the order of several thousand wavenumbers in (Ga,Mn)As, that Mn dopoing does not lead to substantial renormalization of the GaAs host VB. This result on field effect devices is in accord with the analysis of resonant tunneling spectroscopy experiments in (Ga,Mn)As layers as well~\cite{Ohya2011}. It is worth noting that the maximum $m_\mathrm{{opt}}$ such that $x_\mathrm{{eff}} \leq x$ for all the samples in Fig.~\ref{mega} is $\sim$0.8 $m_e$. Using this maximum value of $m_\mathrm{{opt}}$ does not change the qualitative behavior of the data in Fig.~\ref{mega}e, which we discuss in the following paragraphs.

Having obtained an estimate of $x_\mathrm{{eff}}$, we plot $T_C/x_\mathrm{{eff}}$ versus $N_{\mathrm{eff}}$/M$_\mathrm{{eff}}$, where M$_\mathrm{{eff}}$ is the density of effective Mn moments (M$_\mathrm{{eff}}=\frac{4}{a^3}x_\mathrm{{eff}}$) in Fig.~\ref{mega}d. The value of $N_{\mathrm{eff}}$/M$_\mathrm{{eff}}$ gives an approximation of the relative degree of compensation of each film. For $N_{\mathrm{eff}}$/M$_\mathrm{{eff}}<$1.25 $m_e^{-1}$, these data show a monotonic increase in $T_C/x_\mathrm{{eff}}$ for increasing $N_{\mathrm{eff}}$/M$_\mathrm{{eff}}$. This general trend is expected in both a $p$-$d$ exchange scenario~\cite{Jungwirth2005}, as well a double-exchange scenario~\cite{Sato2010} for Ga$_{1-x}$Mn$_x$As. For $N_{\mathrm{eff}}$/M$_\mathrm{{eff}}>$1.25 $m_e^{-1}$, the data seem to be maximized in a range of roughly 1.3 $m_e^{-1}<N_{\mathrm{eff}}$/M$_\mathrm{{eff}}<$1.6 $m_e^{-1}$. We note the film with the highest $T_C$ (182 K) falls within this latter range, while the film with the second highest $T_C$ (179 K) falls within the former range. Continuing to higher values of $N_{\mathrm{eff}}$/M$_\mathrm{{eff}}$, the data then show either a plateaued or decreasing trend, which is not easily discerned given the error bars. However, the $x$=0.015 samples show a significant overall decrease in $T_C/x_\mathrm{{eff}}$, displaying a suppression of over 50\%, while having among the largest values of $N_{\mathrm{eff}}$/M$_\mathrm{{eff}}$.

The theoretical curve of $T_C/x_\mathrm{{eff}}$ versus $N_{\mathrm{eff}}$/M$_\mathrm{{eff}}$ calculated in Ref.~\cite{Jungwirth2005} is displayed in Fig.~\ref{mega}e. This curve is based on an electronic structure, calculated using a microscopic tight-binding approximation, in which the valence and impurity bands of Ga$_{1-x}$Mn$_x$As have completely merged. The ferromagnetic properties are thus calculated according to a semiphenomenological $p$-$d$ exchange mean-field approximation. The key theoretical prediction of Ref.~\cite{Jungwirth2005}, is a systematic, monotonically increasing trend in $T_C/x_\mathrm{{eff}}$, with a corollary that the highest $T_C$ will occur for films with close to zero compensation. Furthermore, to the best of our knowledge, all VB $p$-$d$ exchange descriptions predict a monotonic increase of $T_C$ with both $p$ and $x_\mathrm{{eff}}$. 

According to our IR spectral weight analysis, the key prediction made in the $p$-$d$ exchange framework (monotonic increase of $T_C/x_\mathrm{{eff}}$) is not in agreement with the data. Non-monotonic trends of $T_C/x_\mathrm{{eff}}$ have also been reported in Ref.~\cite{Dobrowolska2012} using a simultaneous combination of channeling Rutherford backscattering (c-RBS) and channeling particle-induced X-ray emission (c-PIXE)~\cite{Dobo2012}. In contrast, high-field Hall measurements of the carrier density do not show the non-monotonic behavior seen in Fig.~\ref{mega}d~\cite{Jungwirth2005}. The experimental data of Refs.~\cite{Dobrowolska2012} and ~\cite{Jungwirth2005} have both been disputed~\cite{Edwards2012, Dobo2012}. With this noted disagreement in the literature, we emphasize that our experiments are done using a technique entirely different from any of those above. Furthermore, our IR method is a contactless probe, and the sum-rule analysis is based on model-independent arguments that are rooted in the causality of the electromagnetic response~\cite{Basov2011}. We reiterate that the highest $T_C$ films (182 K and 179 K) displayed in Fig.~\ref{mega} (filled gray square and circle, respectively) have spectral weight that is well below that expected for the uncompensated case, which is contradictory to the predictions of the $p$-$d$ exchange mean-field approximation. Therefore, we argue that the ferromagnetic properties of Ga$_{1-x}$Mn$_x$As require description beyond Zener's $p$-$d$ exchange model.



To our knowledge, there is no microscopic theory that has presented a quantitative description of the non-monotonic behavior of Fig.~\ref{mega}d. It was argued in Ref.~\cite{Dobrowolska2012} that $T_C$ is controlled by the location of $E_F$ in a Mn-induced IB. From the basis of tight-binding Anderson calculations, however, IB models used to describe earlier experiments~\cite{Burch2006, Stone2008, Ando2008, Tang2008} fail to demonstrate how the IB could remain sufficiently narrow as to avoid overlap with the VB at the doping levels necessary to initiate FM~\cite{Masek2010}. A strictly detached IB, however, is not theoretically necessary for the qualitative behavior seen in Fig.~\ref{mega}d. First principle investigations of the electronic structure and magnetism find Ga$_{1-x}$Mn$_x$As to be an intermediate case, with contributions to ferromagnetism coming from both $p$-$d$-like and double exchange-like mechanisms. The calculated DOS of this later study finds significant merging of the IB and VB, yet still with a large amplitude of $d$ states at $E_F$, $and$ a corresponding suppression of $T_C$ as a function of increased hole concentration~\cite{Sato2003}. These same techniques when applied to other dilute magnetic semiconductors show the dome-like behavior of $T_C$ typical of a double exchange FM mechanism becomes further pronounced in cases where the separation of the IB and VB are more distinct, e.g. Ga$_{1-x}$Mn$_x$N. 

The non-monotonic nature of the data of Fig.~\ref{mega}d shows that the FM mechanism of Ga$_{1-x}$Mn$_x$As has a strong double-exchange component. This fact highlights the important role played by impurity states and localization at $E_F$.  A schematic representation of the DOS consistent with our data is shown in the bottom panel of Fig.~\ref{mega}a. Although this schematic lacks subtleties that may vary with parameters such as the Mn concentration and compensation in real materials, a generic optical conductivity lineshape (bottom right panel of Fig.~\ref{mega}a) has been observed in all FM Ga$_{1-x}$Mn$_x$As samples of the studies represented in Fig.~\ref{mega}. This latter fact implies spectroscopic features and carrier dynamics in FM Ga$_{1-x}$Mn$_x$As films reported in the literature originate from the same basic physical picture, despite variations in the growth and film preparation procedures across these works. Furthermore, detailed spectroscopic probes of the IMT in GaMnAs reveal that the generic picture of the DOS in bottom panel of Fig.~\ref{mega}a holds over a wide range of Mn concentrations, persisting over an order of magnitude of Mn doping beyond the onset of conduction in FM Ga$_{1-x}$Mn$_x$As~\cite{Chapler2011}. In Sec.~\ref{intragap}, we expand upon the earlier IR data and further demonstrate that the generic spectral features oberved in Ga$_{1-x}$Mn$_x$As persist over a wide parameter space, and are consistent with an IB scenario.

\begin{table}[]
\centering
\caption{Reference number, nominal doping, film thickness, and ferromagnetic transition temperature for all samples displayed in Fig.~\ref{mega}. ``*'' indicates samples that have been annealed.}
\begin{tabular}{c ||c | c | c } 
\hline 
Reference & nominal doping (\%) & thickness (nm) & T$_C$ (K)\\ 
\hline\hline
\multirow{5}{*}{Ref.~\cite{Singley2002}} & 2.8 & 500 &  30  \\ 
 & 4.0 & 500   & 45  \\
 & 5.2 & 500   & 70  \\
 & 6.6 & 500   & 70  \\
 & 7.9 & 500   & 70  \\
\hline
\multirow{4}{*}{Ref.~\cite{Burch2006}} & 5.2 & 40   & 80  \\ 
 & 5.2* & 40   & 120  \\
 & 7.3 & 40   & 80  \\
 & 7.3* & 40   & 140  \\
\hline
\multirow{6}{*}{Ref.~\cite{Jungwirth2010}} & 2.0* & 20   & 47  \\ 
 & 3.0* & 20  & 77  \\
 & 5.2* & 20   & 132  \\
 & 7.0* & 20   & 159  \\
 & 9.0* & 20   & 179  \\
 & 14* & 20   & 182  \\
\hline
\multirow{8}{*}{This work} & 0.5 & 200   & $<$ 4  \\ 
 & 0.75 & 200   & $<$ 4  \\
 & 1.5 & 200   & 21  \\
 & 1.5* & 200   & 21  \\
 & 3.0 & 200   & 42  \\
 & 3.0* & 200   & 60  \\
 & 9.2* & 100   & 130  \\
 & 16 & 100   & 120  \\
\hline \hline
\end{tabular}

\label{thickness}
\end{table}

\section{Samples and experimental methods}
\label{wtf}
\subsection{Sample growth and characteristics}
\label{growth}
All films in this study were prepared via MBE, on semi-insulating (001) GaAs substrates. The films were prepared using the non-rotated, low-temperature growth techniques reported in Refs.\cite{Myers2006, Mack2008}. Since this study involved several doping regimes, details of the growth vary between samples. Specific details on the film preparation of all samples can be found in Table~\ref{props}. In the Ga$_{1-x}$Mn$_x$As samples, the Mn dopant concentration was determined by growth rate calibrations of MnAs and GaAs RHEED oscillations. For the Ga$_{0.991}$Be$_{0.009}$As sample, the Be concentration was determined from room temperature Hall-effect data.

The non-rotated technique utilizes a geometry in which the MBE system provides a continuous variation in the As:Ga ratio in one direction along the wafer. This yields precise control over the density of As$_{\mathrm{Ga}}$, while holding Mn (or Be) flux approximately constant~\cite{Myers2006}. Local transport studies along the As:Ga gradient demonstrate that increasing the As$_{\mathrm{Ga}}$ density reduces the hole density $p$, mobility and dc conductivity $\sigma_{dc}$ in Ga$_{1-x}$Mn$_x$As~\cite{Myers2006}. Additionally, As$_{\mathrm{Ga}}$ also affects $T_C$, magnetization, and magnetic hysteresis curves in FM samples~\cite{Myers2006, Mack2008}. These studies show that the location of minimum As$_{\mathrm{Ga}}$ density, defined by the location of maximum $p$ found along the As:Ga gradient, also corresponds to the region where all of the above properties are optimized. We note, detailed studies of Ref.~\cite{Myers2006} have concluded full dopant incorporation is not achieved in the As-deficient regions. In the ``optimized'' and As-rich regime, however, the data indicate relatively constant Mn concentration (within 0.1\% variation) along the As:Ga gradient~\cite{Myers2006}. In the heavily alloyed samples ($x$=0.092* and 0.16 Mn (* indicates the sample was annealed)), this latter statement is only true for a small region near the optimal location~\cite{Mack2008}.

Ga$_{1-x}$Mn$_x$As samples of \emph{x}=0.005, 0.0075 were found to be paramagnetic (PM) at all locations along the films. The other more heavily doped Ga$_{1-x}$Mn$_x$As films were found to be FM, with $T_C$ in the region of minimized compensating defects (referred to as MA, with this naming convention explained in Sec.~\ref{expdata} below) listed in Table~\ref{props}. The $x$=0.015*, 0.03*, and 0.092* Ga$_{1-x}$Mn$_x$As samples were also subjected to low-temperature annealing. The $x$=0.015 and 0.015* Ga$_{1-x}$Mn$_x$As film are the same sample pre- and post-annealing; the same is true for the $x$=0.03 and 0.03* Ga$_{1-x}$Mn$_x$As film. In the $x$=0.015* and 0.03* samples, $T_C$ after annealing was determined by the temperature of maximum low-frequency resistivity ($\rho_{\mathrm{IR}}$=1/$\sigma_1$(40 cm$^{-1}$))~\cite{Chapler2011}(see Fig.~\ref{temp}i). For all other samples, including $x$=0.015 and $x$=0.03 before annealing, $T_C$ was measured by SQUID magnetometry.

\begin{table}[]
\centering
\caption{Sample properties of all the films measured in this study including the dopant concentration, film thickness, growth temperature, and annealing temperature (where applicable). ``*'' indicates that samples have been annealed.}
\begin{tabular}{c c || c | c | c} 
\hline 
dopant & \% & thickness (nm) & growth ($^{\circ}{\rm C}$) & annealing ($^{\circ}{\rm C}$)\\ 
\hline\hline
\multirow{8}{*}{Mn} & 0.5 & 200 & 250  & -  \\ 
 & 0.75 & 200 & 250  & -    \\
 & 1.5 & 200 & 250  & -   \\
 & 1.5* & 200 & 250  & 220   \\
 & 3.0 & 200 & 220  & -  \\
 & 3.0* & 200 & 220  & 200   \\
 & 9.2* & 100 & 200  & 180   \\
 & 16 & 100 & 150  & -   \\
\hline
Be & 0.9 & 100 & 250 & -   \\
\hline \hline
\end{tabular}

\label{props}
\end{table}

\subsection{Experimental methods}
\label{expdata}

\begin{figure}[]
\centering
\includegraphics[width=86mm]{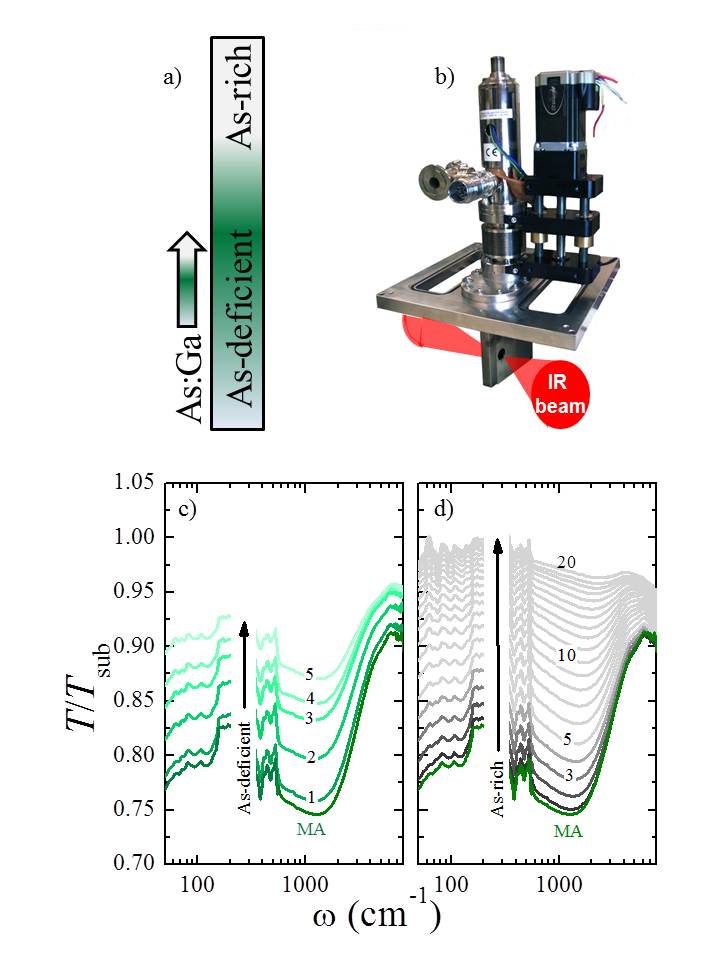}
\caption{a) Schematic of our samples with As:Ga gradient. b) Photograph of our braod-band transmission microscope. c and d Room temperature transmission through the \emph{x}=0.005 Ga$_{1-x}$Mn$_x$As sample, normalized to transmission of the GaAs substrate. Panel c shows transmission data taken at 1 mm increments into the As-deficient region of the sample. Panel d shows transmission spectra taken at 1 mm increments into the As-rich region. Numeric labels indicate the distance in mm from the MA location. The dark green curve displayed in both panels is the transmission spectrum of the MA position.}
\label{rawT}
\end{figure}

In order to characterize the effects of the As:Ga gradient on the optical properties, we developed a broad-band (far-IR to near-ultraviolet) microscope compatible with low-temperature ($\sim$20 K) operation (Fig.~\ref{rawT}b). Our experimental set-up incorporates a 4x beam condenser and linear translation stage in a Fourier transform IR spectrometer to obtain frequency dependent transmission spectra. The spatial resolution of the apparatus is below 1 mm, which is appropriate for the series of films investigated here. This assertion was validated through direct mid-IR transmission microscopy experiments, with the IR-beam diameter focused to less than 100 $\mu$m at the sample, taking measurements in 1 mm increments along the As:Ga gradient. The micro-transmission data was supplemented with room temperature micro-ellipsometry measurements for several films ($x$=0.005, 0.0075, 0.015, 0.03, 0.16 Mn and $x$=0.009 Be). The ellipsometry data span a frequency range of 6,000 cm$^{-1}\leq \omega \leq$ 40,000 cm$^{-1}$, with spatial resolution $\sim$150 $\mu$m. 

Characteristic transmission data along the As:Ga gradient are displayed in Figs.~\ref{rawT}c and d. The figures show room temperature transmission through the \emph{x}=0.005 Ga$_{1-x}$Mn$_x$As sample, normalized to that of the GaAs substrate. We identify the location of maximum $p$, as the position corresponding the maximum absorption along the sample, and thus refer to this location as MA. A more rigorous and quantitative determination of the location of maximum $p$ is completed via optical sum rules (Eq.~\ref{drude}), as detailed later in the manuscript (Sec.~\ref{sectionA}). With this definition of the MA point, the spectra can be divided into two regimes: As-rich, and As-deficient. Spectra in the As-deficient regime show a rapid increase in overall transmission as the IR probe moves deeper into the deficient regime. Detailed studies reported in Ref.~\cite{Myers2006} have concluded full dopant incorporation is not achieved in the As-deficient regions. A reduction in dopant concentration is consistent with the decreased absorption observed. Thus, the IR studies in this work focus on the As-rich and MA positions of the samples. In these latter locations, a systematic increase in the overall transmission is observed as the IR beam is moved from MA deeper into the As-rich region. The increase of transmission in the As-rich regime is consistent with reduction in $p$ due to As$_{\mathrm{Ga}}$ compensation.

\begin{figure}[]
\centering
\includegraphics[width=86mm]{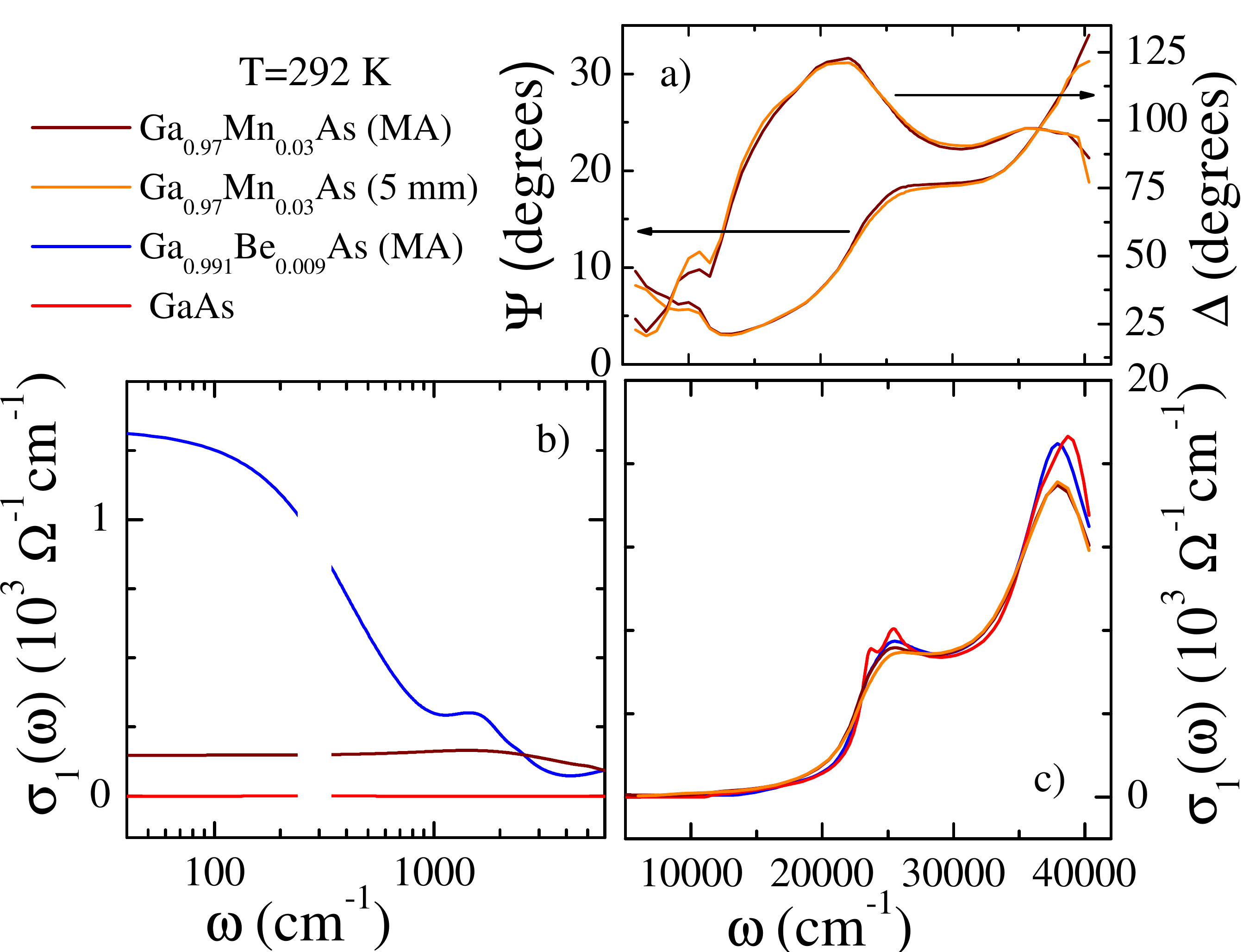}
\caption{Panel a shows the ellipsometric angles $\Psi$ and $\Delta$ for the $x$=0.03 Ga$_{1-x}$Mn$_x$As for MA and 5 mm into the As-rich region. The angle of incidence for data in panel a was 75$^{\circ}$. The $\sigma_1(\omega)$ spectra of the $x$=0.03 Ga$_{1-x}$Mn$_x$As film, the Ga$_{0.991}$Be$_{0.009}$As film, and pristine GaAs in the intragap region are shown in panel b, and extending beyond the GaAs band gap in panel c. Panel c additionally displays spectrum of the $x$=0.03 Ga$_{1-x}$Mn$_x$As 5mm into the As-rich region. The $\sigma_1(\omega)$ data are obtained by a simultaneous fitting of transmission and ellipsometric data, as described in the text.
}
\label{ellips}
\end{figure}

Spectroscopic data at frequencies beyond the fundamental GaAs band gap were obtained via ellipsometry measurements. Ellipsometry measures the ratio of the complex reflectivity coefficient ($r$) of $s$- and $p$-polarized light. The experimental observables are the ellipsometric angles $\Psi$ and $\Delta$ according to,

\begin{equation}
{\frac{r_p}{r_s}=\tan(\Psi)e^{i\Delta}},
\label{ellipseq}
\end{equation}

\noindent where the subscripts $p$ and $s$ denote the incident polarization. Fig.~\ref{ellips}a shows representative data of our micro-ellipsometry experiments for the $x$=0.03 Ga$_{1-x}$Mn$_x$As film at the MA position.

To develop a quantitative understanding of the optical data, it is instructive to express the our observables in terms of the complex IR conductivity spectrum $\sigma(\omega$)=$\sigma_1(\omega$)+i$\sigma_2(\omega)$. We extract the complex conductivity spectrum from our optical data via multi-oscillator modeling. For this model, we use only Kramers-Kronig (KK) consistent oscillators and perform a simultaneous fitting of both the transmission and ellipsometric data. In this way, we construct a KK consistent model of $\sigma(\omega$) over the entire experimental range. Modeling the data over such a large experimental frequency range assures a high degree of confidence in the uniqueness and accuracy of each experimental fit~\cite{Burch2004, Kuzmenko2005}. We focus on the real part of the conductivity spectrum, $\sigma_1(\omega)$, describing the dissipative processes in the system. Fig.~\ref{ellips}b and c display representative plots illustrating $\sigma_1(\omega)$ extracted by the simultaneous  fitting of the room temperature transmission spectrum and ellipsometric data as described above. The data are displayed for the Ga$_{0.97}$Mn$_{0.03}$As and Ga$_{0.991}$Be$_{0.009}$As samples at MA, both in the intragap region (Fig.~\ref{ellips}b), as well as frequencies extending above the fundamental GaAs band gap (Fig.~\ref{ellips}c). Further discussion of the spectroscopic features in the intragap regime are found in Sec.~\ref{intragap}. 

\section{Intragap response of $p$-doped Galium Arsenide}
\label{intragap}
\subsection{Carrier density gradient}
\label{sectionA}

\begin{figure}[]
\centering
\includegraphics[width=86mm]{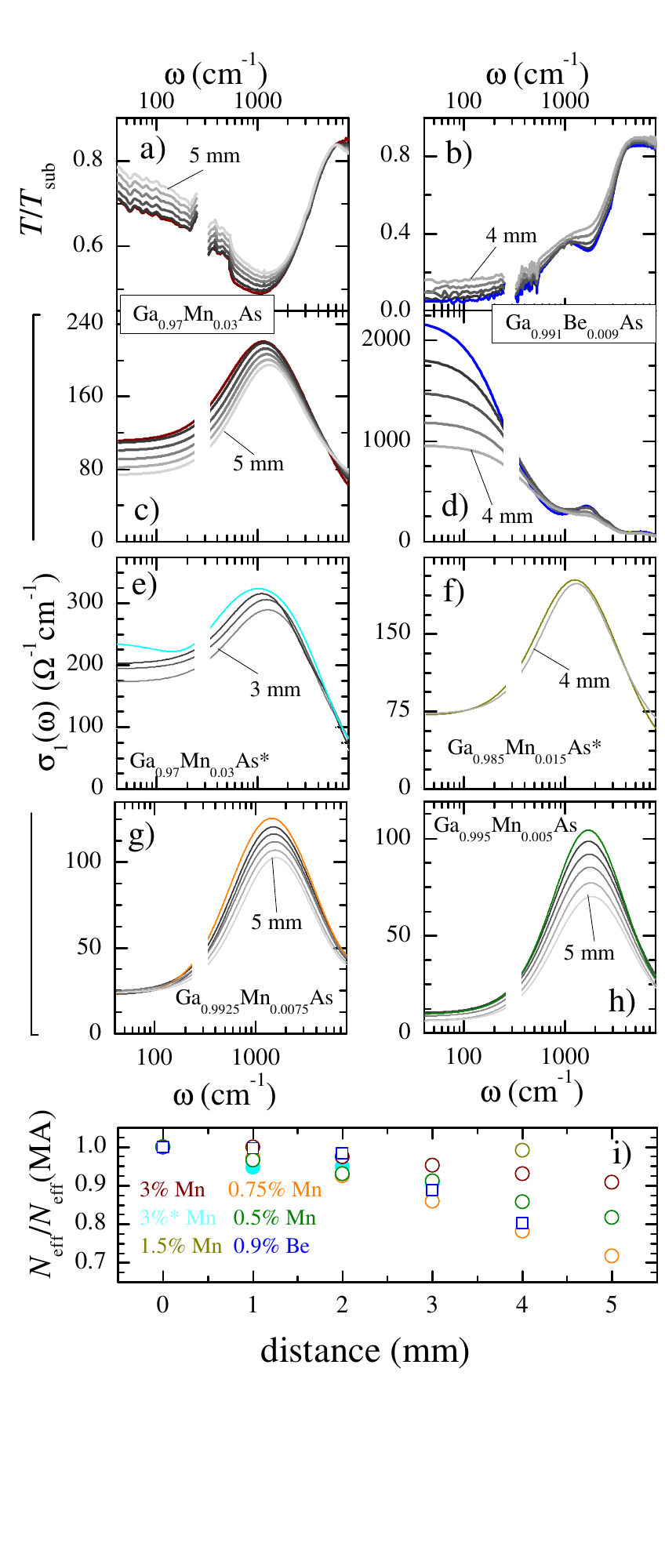}
\caption{Panels a and b show transmission data normalized to the GaAs substrate taken at 20 K for the $x$=0.03 Ga$_{1-x}$Mn$_x$As and Ga$_{0.991}$Be$_{0.009}$As samples, respectively. Panels c and d show the corresponding $\sigma_1(\omega)$ spectra for the $x$=0.03 Ga$_{1-x}$Mn$_x$As and Ga$_{0.991}$Be$_{0.009}$As samples, respectively. The data are displayed at the MA point, which appear in color, and at 1 mm increments into As-rich region appearing in gray scale (increasingly lighter with distance from MA). The breaks in the data near 300 cm$^{-1}$ are due to a GaAs phonon. The low-temperature $\sigma_1(\omega)$ spectra for the $x$=0.005, 0.0075, 0.015, 0.015* and 0.03$^*$ Ga$_{1-x}$Mn$_x$As films at MA, and 1 mm increments into the As-rich region are shown in panels e--h. Panel i shows the spectral weight $N_\mathrm{{eff}}$ (Eq.~\ref{drude}) as a function of distance from the MA point, normalized to $N_\mathrm{{eff}}$ at MA for the data displayed in panels c-h.
}
\label{allgrad}
\end{figure}

Figs.~\ref{allgrad}a--d show representative transmission spectra and corresponding $\sigma_1(\omega)$ for the Ga$_{0.97}$Mn$_{0.03}$As and Ga$_{0.991}$Be$_{0.009}$As samples. These panels highlight the equivalence of the location along the As:Ga gradient of maximum absorption in the transmission spectra, to that of the maximum in overall $\sigma_1(\omega)$. Therefore the MA location can be determined by the position along the As:Ga gradient that displays the maximum spectral weight (Eq.~\ref{drude}). Figs.~\ref{allgrad}e--h show the intragap $\sigma_1(\omega)$ data of the other gradient samples measured in this study. Data along the As:Ga gradient of the heavily alloyed samples ($x$=0.09 and 0.16 Ga$_{1-x}$Mn$_x$As) are not reported because it has been determined that quality surfaces and interfaces are only found in a narrow region near MA~\cite{Mack2008}. The effect of the As:Ga gradient in terms of the spatial location is quantified in Fig.~\ref{allgrad}i by the spectral weight $N_\mathrm{{eff}}$, normalized to $N_\mathrm{{eff}}$ at MA for each sample (absolute scale for $N_\mathrm{{eff}}$ can be read off of Fig.~\ref{neffw0}c) . Fig.~\ref{allgrad}i demonstrates suppression of $N_{\mathrm{eff}}$ as As content increases past the optimal concentration to be a generic result for the $p$-doped GaAs samples of this study.

Figs.~\ref{allgrad}a--h demonstrate that the qualitative features of the spectra, although different for Be and Mn-doped samples, are not radically modified by additional As$_{\mathrm{Ga}}$ compensation. Ga$_{1-x}$Mn$_x$As $\sigma_1(\omega)$ spectra, both at MA and several mm into the As-rich regime, reveal a lineshape consisting of a broad mid-IR resonance and relatively flat conductivity in the far IR. In the Ga$_{1-x}$Be$_x$As data of in Fig.~\ref{allgrad}, the spectra at MA and in the As-rich region are dominated by a pronounced low-frequency response. This latter peak can be modeled semi-classically as a Drude peak, given by: 

\begin{equation}
\sigma_1(\omega)=\frac{\sigma_{dc}}{1+\tau^2\omega^2}.
\label{DrudeEq}
\end{equation}

\noindent The Drude peak is characteristic of a free carrier response in a metal or degenerate semiconductor, with the amplitude equal to the dc conductivity $\sigma_{dc}$ and the width of the peak quantifying the free carrier scattering rate $1/\tau$. As can be seen in the figure, the increase of As$_{\mathrm{Ga}}$ in the As-rich region of Ga$_{1-x}$Be$_x$As suppresses the Drude conductivity and broadens the Drude width. A narrow mid-IR resonance is also suppressed and broadened in the As-rich regions.

\subsection{Doping dependence}
\label{doping}

\begin{figure}[]
\centering
\includegraphics[width=75mm]{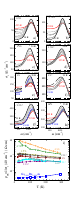}
\caption{Panels a--h display the temperature dependence of the $\sigma_1(\omega)$ spectra (at MA) for all of our Ga$_{1-x}$Mn$_x$As films. Room temperature spectra are shown in red, the lowest temperature spectra in black, and the minima in conductivity near T=T$_C$ are in blue. All other temperatures are shown in gray for clarity. The breaks in the data near 300 cm$^{-1}$ are due to a GaAs phonon. Panel i plots the temperature dependence of $\rho_{IR}$ (defined in the text) for all the samples at the MA position.
}
\label{temp}
\end{figure}

The temperature dependence of $\sigma_1(\omega)$ for all the Ga$_{1-x}$Mn$_x$As samples in this study at MA are shown in Fig.~\ref{temp}. We observe a qualitatively similar lineshape at all Mn-dopings $x$ (see Fig.~\ref{mega}a). The far-IR conductivity, attributed to the itinerant carrier response, is nearly completely suppressed at the lowest temperatures in the PM samples ($x<$1\%). This vanishing conductivity supports the notion that these samples reside on the insulating side of the Ga$_{1-x}$Mn$_x$As insulator-to-metal transition (IMT). In these insulating samples, the observed broad mid-IR resonance in the vicinity of Mn-acceptor binding energy has a natural assignment of VB to IB transitions. 

The persistence of the far-IR conductivity down to the lowest temperatures in Ga$_{1-x}$Mn$_x$As samples of $x>$0.0075 reveals films in this dopant regime to be beyond the onset of conduction. However, the onset of conduction in Ga$_{1-x}$Mn$_x$As remains distinct from genuine metallicity, as can be seen in Fig~\ref{temp}i. In this panel, we plot the temperature dependence of the ``infrared resistivity'' ($\rho_{IR}$=1/$\sigma_1$(40 cm$^{-1}$)). The data in the dilute Mn-doped samples ($x$=0.005, 0.0075) show the systematic increase in $\rho_{IR}$ expected in the case of thermally activated transport. The insulating behavior of these samples down to the lowest temperatures measured establishes the Ga$_{1-x}$Mn$_{x}$As films in this doping regime as below the onset of conduction. 

The onset of conduction is marked by the finite $\sigma_{dc}$ in the limit of $\omega$, $T\rightarrow$0, exhibited by the Ga$_{1-x}$Mn$_{x}$As films of $x>$0.0075. Nevertheless, the $\rho_{IR}$ of the films in this latter dopant regime still display signs of activated transport above $T_C$. The onset of ferromagnetism radically alters the temperature dependence of $\sigma_1(\omega)$ and $\rho_{IR}$, as below $T_C$ the activated character is reversed. An anomalous increase of the low energy spectral weight with the development of magnetization has also been demonstrated in earlier work on Ga$_{1-x}$Mn$_{x}$As~\cite{Hirakawa2002, Singley2002, Chapler2011}, In$_{1-x}$Mn$_{x}$As~\cite{Hirakawa2001}, as well as a canonical double-exchange material La$_{1-x}$Sr$_x$MnO$_3$~\cite{Okimoto1995, Basov2011}. This ``mixed'' behavior of insulating and metallic trends underscores the unconventional nature of conduction in Mn-doped GaAs beyond the IMT boundary, and serves as a signature of the deep bond between magnetism and the carrier dynamics. Moreover, it establishes the distinction between the onset of conduction and genuine metallicity, the latter of which we now describe below for Ga$_{0.991}$Be$_{0.009}$As.

The room temperature spectrum of Ga$_{0.991}$Be$_{0.009}$As (blue curve in Fig.~\ref{ellips}c) reveals a pronounced Drude peak, and a narrow mid-IR resonance. Upon cooling to 25 K, the Drude peak sharpens and increases in amplitude, typical of genuinely metallic behavior, as does the mid-IR resonance (compare blue curves of Figs.~\ref{ellips}b and~\ref{allgrad}d). The temperature dependence of $\rho_{IR}$ for this film reveals a metallic trend throughout the entire measured temperature range (Fig~\ref{temp}c). The pronounced Drude peak and metallic temperature dependence establish the genuine metallicity of this film, in contrast with the mixed behavior of Ga$_{1-x}$Mn$_{x}$As beyond the onset of conduction. Furthermore, much weaker far-IR spectral weight is found in Mn-doped samples than that of the Be-doped film~\cite{Chapler2011}. 

The carrier density can be readily measured by the Hall effect in the non-magnetic Ga$_{0.991}$Be$_{0.009}$As system. From this fact, and coupled with the prominant Drude peak the effective carrier mass can be determined from Eq.~\ref{drude}.  This effective mass and was found to be 0.29 $m_e$ at MA for the Ga$_{0.991}$Be$_{0.009}$As sample. This value of the carrier mass is supported by the mass extracted from mobility data of $p$-type GaAs doped with nonmagnetic (Zn, C, or Be) acceptors~\cite{Alberi2008}. Moreover, is in good agreement with the two-band transport mass of light and heavy holes in the GaAs VB of 0.38 $m_e$~\cite{Wiley1970}. The combination of the light effective carrier mass and metallic temperature dependence of the Drude peak are strong evidence that the IR spectra of the Ga$_{0.991}$Be$_{0.009}$As film are representative of the IR response of carriers that reside in extended states of the GaAs host VB.

Decoupling the delocalized Drude-like contributions to the spectra from interband transition contributions limit analysis similar to that above for estimating carrier masses in Ga$_{1-x}$Mn$_{x}$As. However, reasonable estimates reveal the effective mass of carriers in Ga$_{1-x}$Mn$_{x}$As to be significantly larger than that of the metallic Be-doped sample. The large mass of charge carriers in Ga$_{1-x}$Mn$_{x}$As are apparent from the much weaker far-IR spectral weight in Ga$_{1-x}$Mn$_{x}$As samples with similar doping to that of the Be-doped film. We further note that the weak far-IR spectral weight of Ga$_{1-x}$Mn$_x$As samples extends to films with over an order of magnitude higher dopant concentration than the Ga$_{0.991}$Be$_{0.009}$As film. More detailed quantitative analysis of Ga$_{1-x}$Mn$_{x}$As effective masses in IR data can be found in Ref.~\cite{Singley2003, Burch2006}. The relatively large carrier mass is indicative of transport within a narrow band of impurity character. Moreover, the weak far-IR conductivity and coexistence of insulating and metallic trends in the temperature dependence establish that descriptions of electronic conduction in Ga$_{1-x}$Mn$_{x}$As require an emphasis on localization. These latter characteristics are distinct from extended states in the VB, and are in line with proposed theories of IR conductivity of Ga$_{1-x}$Mn$_{x}$As~\cite{Moca2009a, Bouzerar2011a}.

\subsection{Mid-IR peak}
\label{peak}

\begin{figure}[]
\centering
\includegraphics[width=80mm]{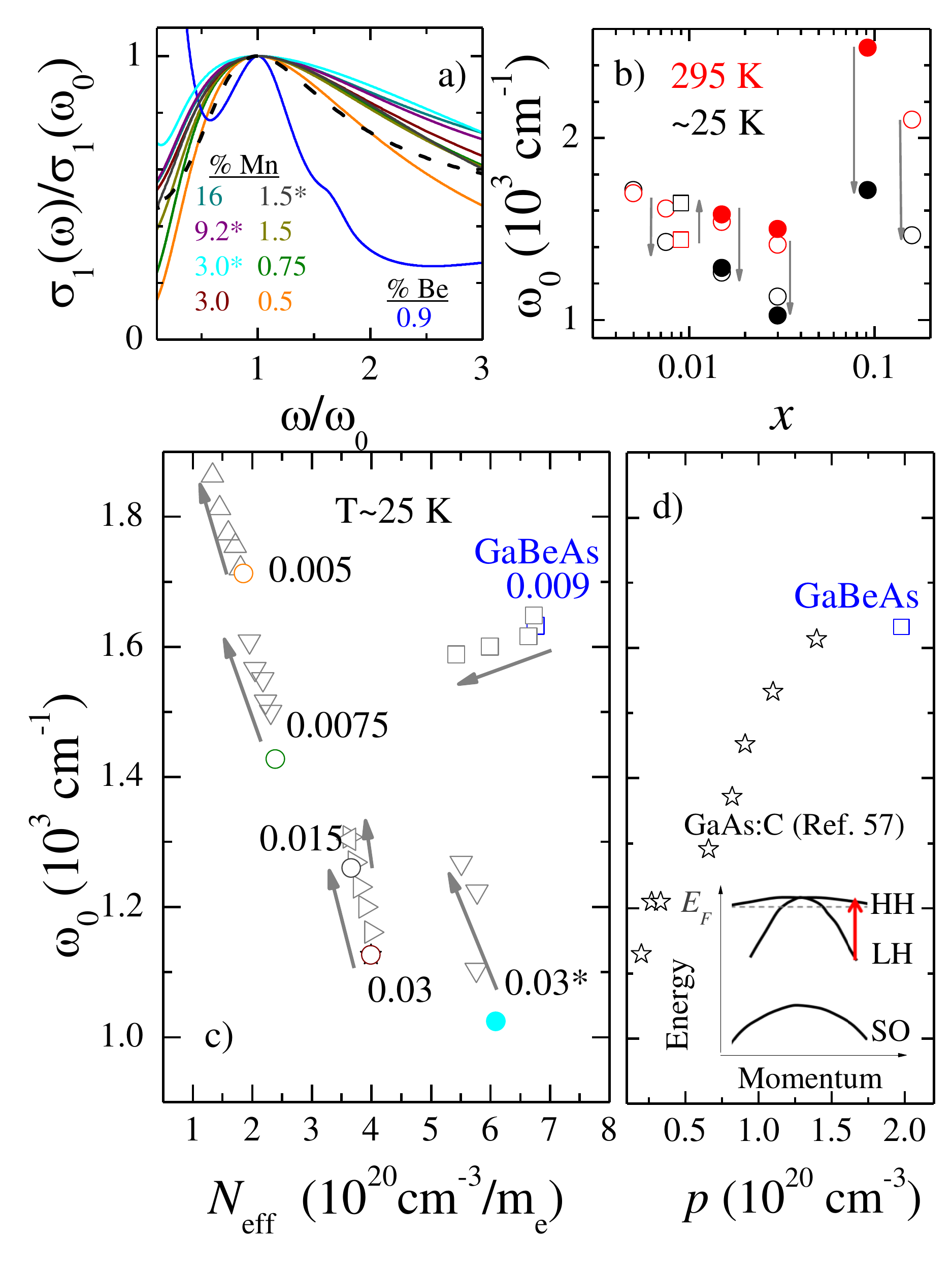}
\caption{Panel a shows $\sigma_1(\omega)$ normalized along the $y$ axis by the value of the conductivity at the peak frequency ($\omega_0$) of the broad mid-IR resonance in the Ga$_{1-x}$Mn$_x$As, and the narrow mid-IR resonance in Ga$_{0.991}$Be$_{0.009}$As film, respectively. Along the $x$ axis, the spectra are normalized by $\omega_0$ of each film. The black dashed line shows the scaled $\sigma_1(\omega)$ of VB to IB transitions in Ga$_{1-x}$Mn$_x$As calculated by the quantum defect method, as described in the text. (b) Location of MIR peak $\omega_0$ as a function of doping $x$. Filled circles in panel a denote annealed Ga$_{1-x}$Mn$_x$As samples, while open circles indicate as-grown Ga$_{1-x}$Mn$_x$As films. The open squares are the data for the Ga$_{0.991}$Be$_{0.009}$As sample. The red data points represent room temperature values, while black points indicate low-temperature data ($\sim$25K). Grey arrows indicate direction of frequency shift upon cooling. (c) Location of $\omega$$_0$ as a function of \emph{$N_{\mathrm{eff}}$} as defined by Eq.~\ref{drude}. The gray arrows indicate direction of increase in As$_{\mathrm{Ga}}$ compensation. The gray points indicate As-rich positions, colored points the MA position, and the filled circle denotes the sample has been annealed. (d) $\omega_0$ as a function of $p$ for the GaAs:C films from Ref.\cite{Songprakob2002}, as well as the MA position of the Ga$_{0.991}$Be$_{0.009}$As film at 20 K. The inset shows the valence band structure of GaAs, with the red arrow indicating excitations attributed to the mid-IR peak in the Ga$_{0.991}$Be$_{0.009}$As data and GaAs:C data of Ref.\cite{Songprakob2002}.}
\label{neffw0}
\end{figure}

In the discussion above, we demonstrated the notable differences in the far-IR behavior of the spectrum of Ga$_{0.991}$Be$_{0.009}$As to that of Ga$_{1-x}$Mn$_x$As. Thus it is interesting to note there is a mid-IR resonance in Ga$_{0.991}$Be$_{0.009}$As with center frequency similar to that of the broad resonance observed in the Ga$_{1-x}$Mn$_x$As spectra. We argue, however, these features stem from different origins. To highlight the difference between the mid-IR feature in Ga$_{0.991}$Be$_{0.009}$As and that of Ga$_{1-x}$Mn$_x$As, we plot ``scaled'' conductivity spectra (of the MA location) in Fig.~\ref{neffw0}a. The figure shows $\sigma_1(\omega)$ normalized along the $y$ axis by the value of the conductivity at the peak frequency ($\omega_0$) of the broad mid-IR resonance in the Ga$_{1-x}$Mn$_x$As, and the narrow mid-IR resonance in Ga$_{0.991}$Be$_{0.009}$As film, respectively. Along the $x$ axis, the spectra are normalized by $\omega_0$ of each film. 

The data in Fig.~\ref{temp} demonstrate that all the Mn-doped samples (both in the insulating regime and those past the onset of conductivity) show very broad, structureless resonances at mid-IR frequencies. Furthermore, upon scaling of these data (Fig.~\ref{neffw0}a), the Mn-doped samples reveal a nearly identical lineshape, barring small non-monotonic differences in the width of the mid-IR resonance. The similar lineshape suggests a similar origin of the mid-IR resonance in both insulating and conducting samples (VB to IB optical excitations).

The persistence of the mid-IR resonance in Ga$_{1-x}$Mn$_x$As samples on the insulating side of the IMT, where an IB is expected for Ga$_{1-x}$Mn$_x$As and non-controversial~\cite{Moriya2003, Jungwirth2007}, to samples well above the onset of conductivity, is key evidence that impurity states maintain a prominent role in FM Ga$_{1-x}$Mn$_x$As. We add to this evidence by calculating the $\sigma_1(\omega)$ lineshape for VB to IB transitions in Ga$_{1-x}$Mn$_x$As according to the quantum defect method for band-to-acceptor transitions~\cite{Bebb1967} (black dashed line in Fig.~\ref{neffw0}a). This model was first applied to absorption in semiconductors by Bebb $et$ $al.$~\cite{Bebb1969}, and later used to model absorption in magnetic semiconductors such as Cd$_{1-x}$Mn$_x$Te~\cite{Huber1987} and Ga$_{1-x}$Mn$_x$As~\cite{Kojima2007}. The details of our calculation can be found in Ref.~\cite{Chapler2011}. Evident in Fig.~\ref{neffw0}a is that the lineshape calculated by the quantum defect method is in good agreement with the lineshape of the mid-IR resonance observed in the Ga$_{1-x}$Mn$_x$As samples. More complex calculations of VB to IB transitions in  Ga$_{1-x}$Mn$_x$As have also been shown to capture key aspects of the experimentally observed mid-IR resonance Refs.~\cite{Moca2009a, Bouzerar2011a}. Thus the agreement of theoretical calculations and the experimental reality support the assignment of the mid-IR resonance of Ga$_{1-x}$Mn$_x$As to VB to IB excitations.

In contrast to the broad resonance observed in the Ga$_{1-x}$Mn$_x$As films, Fig.~\ref{neffw0}a shows the mid-IR resonance observed in Ga$_{0.991}$Be$_{0.009}$As is much narrower, and has a two-peak structure. The second peak appears as a shoulder on the right hand side of the main peak, seen near $\omega/\omega_0$=1.5 in Fig.~\ref{neffw0}a. The frequency position of the two-peak structure is an order of magnitude higher than the Be acceptor binding energy~\cite{Nagai2005}. However, this structure is near the energy expected for intra-VB transitions, in accord with $E_F$ located deep within the VB. The main peak is thus attributed light-hole band (LH) to heavy-hole band (HH) excitations~\cite{Songprakob2002}, while the shoulder is due to excitations from the split-off band (SO)\cite{Braunstein1962} (see Fig.~\ref{neffw0}d). 

The peak frequency $\omega_0$ of the broad mid-IR resonance in Ga$_{1-x}$Mn$_x$As, and narrow mid-IR resonance in Ga$_{1-x}$Be$_x$As, are plotted in Fig.~\ref{neffw0} as a function of doping (Fig.~\ref{neffw0}b) (at MA location only) and $N_{\mathrm{eff}}$ (Fig.~\ref{neffw0}c) (gradient data included). Earlier IR studies of Ga$_{1-x}$Mn$_x$As revealed a systematic red-shift of $\omega_0$ as a function of effective carriers~\cite{Singley2002, Burch2006}. However, other more recent experiments have reported $\omega_0$ to blue-shift with increased Mn-doping~\cite{Jungwirth2010}. We note, data reported in Ref.~\cite{Jungwirth2010} were taken at room-temperature, adding some ambiguity to a direct comparison to the 7 K data of Ref.~\cite{Burch2006}. The difficulty in comparing room temperature and low-temperature data is highlighted in Fig.~\ref{neffw0}b. The figure shows that when Ga$_{1-x}$Mn$_x$As samples are cooled from room temperature to low temperature ($\sim$25 K), $\omega_0$ has a pronounced red-shift. This latter trend holds true for all the Ga$_{1-x}$Mn$_x$As samples except the lowest doped, insulating paramagnetic $x$=0.005 sample, which shows virtually no change in $\omega_0$ upon cooling. The exact opposite trend is seen in Ga$_{1-x}$Be$_x$As, where a small blue-shifts of $\omega_0$ is observed upon cooling from room to low-temperature. 

In our Ga$_{1-x}$Mn$_x$As samples, considering first only the MA location, the data do not show a singular red or blue shift trend of $\omega_0$ over the entire doping range measured. Starting with the dilutely Mn-doped samples, the data in Fig.~\ref{neffw0}b show $\omega_0$ appears to red-shift as the doping is increased, both at room and low-temperature. However, there is a large deviation of the red-shift trend when the doping is extended to Ga$_{0.908}$Mn$_{0.092}$As* and Ga$_{0.84}$Mn$_{0.16}$As. Frequency shifts of $\omega_0$ between Ga$_{1-x}$Mn$_x$As samples in an IB scenario are not dependent just on the nominal doping, but the values of $p$, and $p$/$x$ as well~\cite{Moca2009a, Bouzerar2011a}. These latter facts, coupled with the systematic temperature dependence of $\omega_0$, may indicate why trends in $\omega_0$ versus $x$ are non-universal, and why relative red or blue shifts may be found without these results being contradictory. We note that by changing $p$ while holding $x$ constant, $E_F$ may be tuned without otherwise altering the electronic structure (at least to lowest order approximation). Thus we point out, there are two techniques for changing $p$ while holding $x$ constant from our data: via annealing and via As:Ga gradient. The effects of annealing and of the As:Ga gradient on $\omega_0$ are discussed in the following paragraphs. 

We examine the effect of annealing Ga$_{1-x}$Mn$_x$As on $\omega_0$, focusing on the MA location. The $x$=0.015 sample data show very small change in $N_{\mathrm{eff}}$ after annealing ($\sim$7\% increase). This small change in $N_{\mathrm{eff}}$ is consistent with the assertion of a low degree of compensation in samples in doping regimes below a few atomic percent. Additionally, after annealing the $x$=0.015 sample shows no significant change in $\omega_0$. However, $N_{\mathrm{eff}}$ in the $x$=0.03 Mn-doped sample was significantly enhanced after annealing ($\sim$54\% increase). If we compare only the room-temperature spectra of this latter film, $\omega_0$ in the $x$=0.03 Mn-doped sample exhibits a blue-shift after annealing. This blue-shift is consistent with findings on room-temperature spectra of samples pre- and post-annealing in Ref.~\cite{Jungwirth2010}. However, if instead only the low-temperature data of this film is compared, the $\omega_0$ data of the $x$=0.03 Mn-doped sample red-shifts after annealing. These low-temperature results are consistent with the 7 K data in Ref.~\cite{Burch2006}. Thus the trends in annealing induced shifts of $\omega_0$ in the $x$=0.03 Ga$_{1-x}$Mn$_x$As sample are opposite when room-temperature data is compared as to when low-temperature data is compared. This latter fact underscores the difficulty in comparing trends in $\omega_0$ taken at room temperature to that of low temperatures. Furthermore, it emphasizes that the results of Refs.~\cite{Burch2006} and~\cite{Jungwirth2010} are not contradictory.

Turning to Fig.~\ref{neffw0}c, in each Ga$_{1-x}$Mn$_x$As sample where $N_{\mathrm{eff}}$ is tuned by As$_{\mathrm{Ga}}$ compensation due to non-rotated growth, a systematic redshift of $\omega_0$ is observed as $N_{\mathrm{eff}}$ is increased. Again, the exact opposite trend is observed in Ga$_{1-x}$Be$_x$As, where $\omega_0$ blue shifts as $N_{\mathrm{eff}}$ is increased. The blue-shift trend in Ga$_{1-x}$Be$_x$As is consistent with that of GaAs:C, another $p$-doped GaAs material determined to be metallic, with conduction occurring the VB of the host~\cite{Songprakob2002} (see Fig.~\ref{neffw0}c). Unfortunately, similar gradient data is not available on the heavily alloyed Mn-doped samples due to the nature of their growth (Sec.~\ref{wtf}). 

The observed red-shift of $\omega_0$ in Ga$_{1-x}$Mn$_x$As samples upon annealing and As$_{\mathrm{Ga}}$ reduction is all in accord with detailed theoretical calculations of $\sigma_1(\omega)$ of Ga$_{1-x}$Mn$_x$As near the onset of conduction in an ``IB scenario''~\cite{Moca2009a, Bouzerar2011a}. We emphasize that the opposite trend (blue-shift) with respect to carrier density was observed in the ``VB metals'' Ga$_{1-x}$Be$_x$As and GaAs:C, which is consistent with expectations for $p$-doped GaAs with $E_F$ within the VB~\cite{Burch2006}. Although it has been shown that theoretically the screening of disorder could result in redshift of $\omega_0$ as $x$ and $p$ are increased in a VB scenario~\cite{Jungwirth2007}, experimentally, the trends in $\omega_0$ observed are inconsistent with those observed for VB metals. We cannot fully account for the change in trend of $\omega_0$ in heavily alloyed Mn-doped samples, therefore we only speculate that more detailed calculations of the electronic structure of the VB and IB in heavily alloyed samples may be needed. A final note in this regard is that the low-temperature data show $\omega_0$ is never at significantly higher energy than that of the dilutely Mn-doped $x$=0.005 sample. This latter fact, in addition to the near identical lineshape, further suggests that the mid-IR resonance of the heavily alloyed samples is of the same nature of that of the lower doped Ga$_{1-x}$Mn$_x$As films.

\section{Conclusion and outlook}
\label{conclusion}

The totality of our data demonstrate neither the electronic nor magnetic properties of Ga$_{1-x}$Mn$_x$As are in complete accord with expectations of a $p$-$d$ exchange picture. Instead, the data are suggestive that a Mn-induced IB plays an important role in determining the electronic and magnetic properties. Throughout the text we have provided arguments and references to relevant theory that supports an IB description for our data. Furthermore, the conclusion of the presence of a Mn-induced IB is supported by numerous previous studies of the electronic structure in FM Ga$_{1-x}$Mn$_x$As~\cite{Okabayashi2001, Kojima2007, Ando2008, Mayer2010, Burch2008, Burch2006, Ohya2010a, Ohya2011, Chapler2011, Chapler2012}. 

Nonetheless, despite the experimental and theoretical support for an IB picture Ga$_{1-x}$Mn$_x$As mentioned above, the experimental reality of the precise degree of separation between, and ``mixing'' of, impurity states and valence states remains unresolved. Analysis of resonant tunneling experiments has supported a strictly detached IB with remarkably little exchange splitting of the VB~\cite{Ohya2010a, Ohya2011}. The interpretation of these tunneling data, however, has not been universally accepted (see Ref.~\cite{Dietl2011} for alternative interpretation, and Ref.~\cite{Ohya2011B} for a response.) Additionally, it is difficult to reconcile a completely detached IB with the fact that there seems to be no cases where reduction of Mn$_i$ by annealing (or reduction of As$_\mathrm{Ga}$ by non-rotated growth) results in decreased $T_C$ or conductivity. Decreases in these observables are also conspicuously absent when holes are added by electric field-effect in (Ga,Mn)As-based devices~\cite{Sawicki2009}. These facts suggest that there is significant overlap of impurity states and valence states. Some overlap of these states is also supported by spectroscopic scanning tunneling microscopy experiments, which do not find features consistent with a weakly disorder VB or distinct IB picture, but instead highlight the importance of compensation and disorder in Ga$_{1-x}$Mn$_x$As~\cite{Richardella2010}.

\begin{figure*}[]
\centering
\includegraphics[width=180mm]{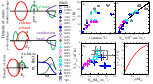}
\caption{a) Schematic diagram of the spin-polarized density of states for the case of double exchange (top) and $p$-$d$ exchange (middle). In the bottom of panel a we show on the left a schematic spin-polarized density of states in the vicinity of $E_F$ for (Ga,Mn)As consistent with our experiments, where gray shaded regions indicate increased localization. The bottom right of panel a is a schematic diagram of the IR conductivity of (Ga,Mn)As, as measured by our experiments. The black curve is the total IR conductivity, whereas the blue curve represents the itinerant carrier response, and the dark green indicates the valence $p$-band to impurity $d$-band interband transition contribution to the spectra. For panels b)-- e) Magenta points are from Ref.~\cite{Singley2002}, cyan from Ref.~\cite{Burch2006}, gray from Ref.~\cite{Jungwirth2010}, and blue from this work (see Table~\ref{thickness} for additional sample details). The data from Ref.~\cite{Singley2002} and Ref.~\cite{Burch2006} were collected at 5--7 K, that of Ref.~\cite{Jungwirth2010} at room temperature, and that of this work at $\sim$25 K. Open points indicate as-grown films, and filled points indicate the film was annealed. b) Ferromagnetic transition temperature $T_C$ as a function of the nominal doping concentration $x$ of Ga$_{1-x}$Mn$_x$As films. c) $T_C$ as a function of the IR spectral weight $N_\mathrm{{eff}}$ (defined in Eq.~\ref{drude}) of Ga$_{1-x}$Mn$_x$As films. The black line is a best-fit power law trendline with exponent of 0.67. d) $T_C$ normalized by the effective Mn concentration $x_{\mathrm{eff}}$ as a function of the IR spectral weight $N_\mathrm{{eff}}$ normalized by the effective Mn moment M$_\mathrm{{eff}}$, as described in the text. e) Theory of Ref.~\cite{Jungwirth2005}, demonstrating the trend expected in $T_C$/$x_{\mathrm{eff}}$ when the FM properties are calculated in a Zener $p$-$d$ exchange framework.
}
\label{mega}
\end{figure*}

We reiterate that the results of first principle investigations show that some overlap of valence and impurity states is not in conflict with our results or conclusions. For instance, the first-principle investigations conclude there is a significant double-exchange contribution, in addition to $p$-$d$ exchange, to the FM state in Ga$_{1-x}$Mn$_x$As for theoretically uncompensated samples~\cite{Sato2010}. The spectroscopic data we report here, and in particular the far-IR response, as well as the non-monotonic trends of $T_C$ of Fig.~\ref{mega}d, point towards an emphasis on localization. Thus it seems a general description of the properties of Ga$_{1-x}$Mn$_x$As requires not just a quantitative picture of the DOS, but also a mobility landscape that quantitatively describes the degree of localization of states in the vicinity of $E_F$. The details of these landscapes will likely be dependent on $x$, the concentration and type of compensation, other sources of disorder, and include possible electron-electron interactions~\cite{Richardella2010}. 

As stated in the introduction, both double exchange and $p$-$d$ exchange may contribute to the resultant FM ground state of Ga$_{1-x}$Mn$_x$As. Understanding how these mechanisms compete and or cooperate as the DOS, localization of states, and general character of the states near $E_F$ evolve as relevant parameters of Ga$_{1-x}$Mn$_x$As are tuned should be vital to a thorough description of this material. Moreover, such an understanding has implications for the broader class of dilute magnetic semiconductors, potentially providing insight into driving future spin-related technology. In any case, our experiments demonstrate that localization, and the role of impurity states cannot be neglected in describing the properties of Ga$_{1-x}$Mn$_x$As samples across a wide range of doping and growth conditions. This includes samples doped below the onset of FM, to those with very high $T_C$ (in the context of Ga$_{1-x}$Mn$_x$As), as well as samples doped below the IMT, to those far beyond the onset of conduction. 

\begin{acknowledgements}
Work at UCSD is supported by the Office of Naval Research. Work at UCSB is supported by the Office of Naval Research and the National Science Foundation. Parts of this work were performed at the Center for Integrated Nanotechnologies, a U.S. Department of Energy, Office of Basic Energy Sciences user facility.
\end{acknowledgements}

\maketitle



\begin{thebibliography}{65}
\expandafter\ifx\csname natexlab\endcsname\relax\def\natexlab#1{#1}\fi
\expandafter\ifx\csname bibnamefont\endcsname\relax
  \def\bibnamefont#1{#1}\fi
\expandafter\ifx\csname bibfnamefont\endcsname\relax
  \def\bibfnamefont#1{#1}\fi
\expandafter\ifx\csname citenamefont\endcsname\relax
  \def\citenamefont#1{#1}\fi
\expandafter\ifx\csname url\endcsname\relax
  \def\url#1{\texttt{#1}}\fi
\expandafter\ifx\csname urlprefix\endcsname\relax\def\urlprefix{URL }\fi
\providecommand{\bibinfo}[2]{#2}
\providecommand{\eprint}[2][]{\url{#2}}

\bibitem[{\citenamefont{Jungwirth et~al.}(2006)\citenamefont{Jungwirth,
  Ma\v{s}ek, Ku\v{c}era, and MacDonald}}]{Jungwirth2006b}
\bibinfo{author}{\bibfnamefont{T.}~\bibnamefont{Jungwirth}},
  \bibinfo{author}{\bibfnamefont{J.}~\bibnamefont{Ma\v{s}ek}},
  \bibinfo{author}{\bibfnamefont{J.}~\bibnamefont{Ku\v{c}era}},
  \bibnamefont{and} \bibinfo{author}{\bibfnamefont{a.~H.}
  \bibnamefont{MacDonald}}, \bibinfo{journal}{Reviews of Modern Physics}
  \textbf{\bibinfo{volume}{78}}, \bibinfo{pages}{809} (\bibinfo{year}{2006}),
  ISSN \bibinfo{issn}{0034-6861},
  \urlprefix\url{http://link.aps.org/doi/10.1103/RevModPhys.78.809}.

\bibitem[{\citenamefont{Burch et~al.}(2008)\citenamefont{Burch, Awschalom, and
  Basov}}]{Burch2008}
\bibinfo{author}{\bibfnamefont{K.~S.} \bibnamefont{Burch}},
  \bibinfo{author}{\bibfnamefont{D.~D.} \bibnamefont{Awschalom}},
  \bibnamefont{and} \bibinfo{author}{\bibfnamefont{D.~N.} \bibnamefont{Basov}},
  \bibinfo{journal}{J. Magn. Magn. Mater.} \textbf{\bibinfo{volume}{320}},
  \bibinfo{pages}{3207} (\bibinfo{year}{2008}), ISSN \bibinfo{issn}{03048853},
  \urlprefix\url{http://apps.isiknowledge.com/full\_record.do?product=UA\&sear%
ch\_mode=GeneralSearch\&qid=3\&SID=3CBnhmAMGPE1DF4mm17\&page=1\&doc=4\&colname%
=WOS}.

\bibitem[{\citenamefont{Sato et~al.}(2010)\citenamefont{Sato, Kudrnovsk\'{y},
  Dederichs, Eriksson, Turek, Sanyal, Bouzerar, Katayama-Yoshida, Dinh,
  Fukushima et~al.}}]{Sato2010}
\bibinfo{author}{\bibfnamefont{K.}~\bibnamefont{Sato}},
  \bibinfo{author}{\bibfnamefont{J.}~\bibnamefont{Kudrnovsk\'{y}}},
  \bibinfo{author}{\bibfnamefont{P.~H.} \bibnamefont{Dederichs}},
  \bibinfo{author}{\bibfnamefont{O.}~\bibnamefont{Eriksson}},
  \bibinfo{author}{\bibfnamefont{I.}~\bibnamefont{Turek}},
  \bibinfo{author}{\bibfnamefont{B.}~\bibnamefont{Sanyal}},
  \bibinfo{author}{\bibfnamefont{G.}~\bibnamefont{Bouzerar}},
  \bibinfo{author}{\bibfnamefont{H.}~\bibnamefont{Katayama-Yoshida}},
  \bibinfo{author}{\bibfnamefont{V.~A.} \bibnamefont{Dinh}},
  \bibinfo{author}{\bibfnamefont{T.}~\bibnamefont{Fukushima}},
  \bibnamefont{et~al.}, \bibinfo{journal}{Rev. Mod. Phys.}
  \textbf{\bibinfo{volume}{82}}, \bibinfo{pages}{1633} (\bibinfo{year}{2010}),
  ISSN \bibinfo{issn}{0034-6861},
  \urlprefix\url{http://link.aps.org/doi/10.1103/RevModPhys.82.1633}.

\bibitem[{\citenamefont{Dietl}(2010)}]{Dietl2010}
\bibinfo{author}{\bibfnamefont{T.}~\bibnamefont{Dietl}}, \bibinfo{journal}{Nat.
  Mater.} \textbf{\bibinfo{volume}{9}}, \bibinfo{pages}{965}
  (\bibinfo{year}{2010}), ISSN \bibinfo{issn}{1476-1122},
  \urlprefix\url{http://www.nature.com/doifinder/10.1038/nmat2898}.

\bibitem[{\citenamefont{Samarth}(2012)}]{Samarth2012a}
\bibinfo{author}{\bibfnamefont{N.}~\bibnamefont{Samarth}},
  \bibinfo{journal}{Nat. Mater.} \textbf{\bibinfo{volume}{11}},
  \bibinfo{pages}{360} (\bibinfo{year}{2012}), ISSN \bibinfo{issn}{1476-1122},
  \urlprefix\url{http://www.nature.com/doifinder/10.1038/nmat3317}.

\bibitem[{\citenamefont{Anderson}(1963)}]{Anderson1963}
\bibinfo{author}{\bibfnamefont{P.~W.} \bibnamefont{Anderson}},
  \bibinfo{journal}{Solid State Physics} \textbf{\bibinfo{volume}{14}},
  \bibinfo{pages}{99} (\bibinfo{year}{1963}), ISSN \bibinfo{issn}{0036-8075}.

\bibitem[{\citenamefont{Sheu et~al.}(2007)\citenamefont{Sheu, Myers, Tang,
  Samarth, Awschalom, Schiffer, and Flatte´}}]{Sheu2007}
\bibinfo{author}{\bibfnamefont{B.~L.} \bibnamefont{Sheu}},
  \bibinfo{author}{\bibfnamefont{R.~C.} \bibnamefont{Myers}},
  \bibinfo{author}{\bibfnamefont{J.-M.} \bibnamefont{Tang}},
  \bibinfo{author}{\bibfnamefont{N.}~\bibnamefont{Samarth}},
  \bibinfo{author}{\bibfnamefont{D.~D.} \bibnamefont{Awschalom}},
  \bibinfo{author}{\bibfnamefont{P.}~\bibnamefont{Schiffer}}, \bibnamefont{and}
  \bibinfo{author}{\bibfnamefont{M.~E.} \bibnamefont{Flatte´}},
  \bibinfo{journal}{Phys. Rev. Lett.} \textbf{\bibinfo{volume}{99}},
  \bibinfo{pages}{227205} (\bibinfo{year}{2007}), ISSN
  \bibinfo{issn}{0031-9007},
  \urlprefix\url{http://link.aps.org/doi/10.1103/PhysRevLett.99.227205}.

\bibitem[{\citenamefont{Dietl et~al.}(2000)\citenamefont{Dietl, Ohno,
  Marsukura, Cibert, and Ferrand}}]{Dietl2000}
\bibinfo{author}{\bibfnamefont{T.}~\bibnamefont{Dietl}},
  \bibinfo{author}{\bibfnamefont{H.}~\bibnamefont{Ohno}},
  \bibinfo{author}{\bibfnamefont{F.}~\bibnamefont{Marsukura}},
  \bibinfo{author}{\bibfnamefont{J.}~\bibnamefont{Cibert}}, \bibnamefont{and}
  \bibinfo{author}{\bibfnamefont{D.}~\bibnamefont{Ferrand}},
  \bibinfo{journal}{Science} \textbf{\bibinfo{volume}{287}},
  \bibinfo{pages}{1019} (\bibinfo{year}{2000}), ISSN \bibinfo{issn}{00368075},
  \urlprefix\url{http://www.sciencemag.org/cgi/doi/10.1126/science.287.5455.10%
19}.

\bibitem[{\citenamefont{Singley et~al.}(2002)\citenamefont{Singley, Kawakami,
  Awschalom, and Basov}}]{Singley2002}
\bibinfo{author}{\bibfnamefont{E.~J.} \bibnamefont{Singley}},
  \bibinfo{author}{\bibfnamefont{R.~K.} \bibnamefont{Kawakami}},
  \bibinfo{author}{\bibfnamefont{D.~D.} \bibnamefont{Awschalom}},
  \bibnamefont{and} \bibinfo{author}{\bibfnamefont{D.~N.} \bibnamefont{Basov}},
  \bibinfo{journal}{Phys. Rev. Lett.} \textbf{\bibinfo{volume}{89}},
  \bibinfo{pages}{097203} (\bibinfo{year}{2002}), ISSN
  \bibinfo{issn}{0031-9007},
  \urlprefix\url{http://link.aps.org/doi/10.1103/PhysRevLett.89.097203}.

\bibitem[{\citenamefont{Burch et~al.}(2006)\citenamefont{Burch, Shrekenhamer,
  Singley, Stephens, Sheu, Kawakami, Schiffer, Samarth, Awschalom, and
  Basov}}]{Burch2006}
\bibinfo{author}{\bibfnamefont{K.~S.} \bibnamefont{Burch}},
  \bibinfo{author}{\bibfnamefont{D.~B.} \bibnamefont{Shrekenhamer}},
  \bibinfo{author}{\bibfnamefont{E.~J.} \bibnamefont{Singley}},
  \bibinfo{author}{\bibfnamefont{J.}~\bibnamefont{Stephens}},
  \bibinfo{author}{\bibfnamefont{B.~L.} \bibnamefont{Sheu}},
  \bibinfo{author}{\bibfnamefont{R.~K.} \bibnamefont{Kawakami}},
  \bibinfo{author}{\bibfnamefont{P.}~\bibnamefont{Schiffer}},
  \bibinfo{author}{\bibfnamefont{N.}~\bibnamefont{Samarth}},
  \bibinfo{author}{\bibfnamefont{D.~D.} \bibnamefont{Awschalom}},
  \bibnamefont{and} \bibinfo{author}{\bibfnamefont{D.~N.} \bibnamefont{Basov}},
  \bibinfo{journal}{Phys. Rev. Lett.} \textbf{\bibinfo{volume}{97}},
  \bibinfo{pages}{087208} (\bibinfo{year}{2006}), ISSN
  \bibinfo{issn}{0031-9007},
  \urlprefix\url{http://link.aps.org/doi/10.1103/PhysRevLett.97.087208}.

\bibitem[{\citenamefont{Jungwirth et~al.}(2010)\citenamefont{Jungwirth,
  Horodysk\'{a}, Tesarov\'{a}, N\v{e}mec, \v{S}ubrt, Mal\'{y}, Ku\v{z}el,
  Kadlec, Ma\v{s}ek, N\v{e}mec et~al.}}]{Jungwirth2010}
\bibinfo{author}{\bibfnamefont{T.}~\bibnamefont{Jungwirth}},
  \bibinfo{author}{\bibfnamefont{P.}~\bibnamefont{Horodysk\'{a}}},
  \bibinfo{author}{\bibfnamefont{N.}~\bibnamefont{Tesarov\'{a}}},
  \bibinfo{author}{\bibfnamefont{P.}~\bibnamefont{N\v{e}mec}},
  \bibinfo{author}{\bibfnamefont{J.}~\bibnamefont{\v{S}ubrt}},
  \bibinfo{author}{\bibfnamefont{P.}~\bibnamefont{Mal\'{y}}},
  \bibinfo{author}{\bibfnamefont{P.}~\bibnamefont{Ku\v{z}el}},
  \bibinfo{author}{\bibfnamefont{C.}~\bibnamefont{Kadlec}},
  \bibinfo{author}{\bibfnamefont{J.}~\bibnamefont{Ma\v{s}ek}},
  \bibinfo{author}{\bibfnamefont{I.}~\bibnamefont{N\v{e}mec}},
  \bibnamefont{et~al.}, \bibinfo{journal}{Phys. Rev. Lett.}
  \textbf{\bibinfo{volume}{105}}, \bibinfo{pages}{227201}
  (\bibinfo{year}{2010}), ISSN \bibinfo{issn}{0031-9007},
  \urlprefix\url{http://link.aps.org/doi/10.1103/PhysRevLett.105.227201}.

\bibitem[{\citenamefont{Jungwirth et~al.}(2005)\citenamefont{Jungwirth, Wang,
  and Edmonds}}]{Jungwirth2005}
\bibinfo{author}{\bibfnamefont{T.}~\bibnamefont{Jungwirth}},
  \bibinfo{author}{\bibfnamefont{K.}~\bibnamefont{Wang}}, \bibnamefont{and}
  \bibinfo{author}{\bibfnamefont{K.}~\bibnamefont{Edmonds}},
  \bibinfo{journal}{Physical Review B} \textbf{\bibinfo{volume}{72}},
  \bibinfo{pages}{165204} (\bibinfo{year}{2005}),
  \urlprefix\url{http://prb.aps.org/abstract/PRB/v72/i16/e165204}.

\bibitem[{\citenamefont{Erwin and Petukhov}(2002)}]{Erwin2002}
\bibinfo{author}{\bibfnamefont{S.~C.} \bibnamefont{Erwin}} \bibnamefont{and}
  \bibinfo{author}{\bibfnamefont{A.~G.} \bibnamefont{Petukhov}},
  \bibinfo{journal}{Phys. Rev. Lett.} \textbf{\bibinfo{volume}{89}},
  \bibinfo{pages}{227201} (\bibinfo{year}{2002}), ISSN
  \bibinfo{issn}{0031-9007},
  \urlprefix\url{http://link.aps.org/doi/10.1103/PhysRevLett.89.227201}.

\bibitem[{\citenamefont{Missous}(1994)}]{Missous1994}
\bibinfo{author}{\bibfnamefont{M.}~\bibnamefont{Missous}}, \bibinfo{journal}{J.
  Appl. Phys.} \textbf{\bibinfo{volume}{75}}, \bibinfo{pages}{3396}
  (\bibinfo{year}{1994}).

\bibitem[{\citenamefont{Ohno et~al.}(1996)\citenamefont{Ohno, Shen, Matsukura,
  Oiwa, Endo, Katsumoto, and Iye}}]{Ohno1996}
\bibinfo{author}{\bibfnamefont{H.}~\bibnamefont{Ohno}},
  \bibinfo{author}{\bibfnamefont{a.}~\bibnamefont{Shen}},
  \bibinfo{author}{\bibfnamefont{F.}~\bibnamefont{Matsukura}},
  \bibinfo{author}{\bibfnamefont{a.}~\bibnamefont{Oiwa}},
  \bibinfo{author}{\bibfnamefont{a.}~\bibnamefont{Endo}},
  \bibinfo{author}{\bibfnamefont{S.}~\bibnamefont{Katsumoto}},
  \bibnamefont{and} \bibinfo{author}{\bibfnamefont{Y.}~\bibnamefont{Iye}},
  \bibinfo{journal}{Appl. Phys. Lett.} \textbf{\bibinfo{volume}{69}},
  \bibinfo{pages}{363} (\bibinfo{year}{1996}), ISSN \bibinfo{issn}{00036951},
  \urlprefix\url{http://link.aip.org/link/APPLAB/v69/i3/p363/s1\&Agg=doi}.

\bibitem[{\citenamefont{Edmonds et~al.}(2004)\citenamefont{Edmonds,
  Boguslawski, Wang, Campion, Novikov, Farley, Gallagher, Foxon, Sawicki,
  Dietl et~al.}}]{Edmonds2004}
\bibinfo{author}{\bibfnamefont{K.}~\bibnamefont{Edmonds}},
  \bibinfo{author}{\bibfnamefont{P.}~\bibnamefont{Boguslawski}},
  \bibinfo{author}{\bibfnamefont{K.}~\bibnamefont{Wang}},
  \bibinfo{author}{\bibfnamefont{R.}~\bibnamefont{Campion}},
  \bibinfo{author}{\bibfnamefont{S.}~\bibnamefont{Novikov}},
  \bibinfo{author}{\bibfnamefont{N.}~\bibnamefont{Farley}},
  \bibinfo{author}{\bibfnamefont{B.}~\bibnamefont{Gallagher}},
  \bibinfo{author}{\bibfnamefont{C.}~\bibnamefont{Foxon}},
  \bibinfo{author}{\bibfnamefont{M.}~\bibnamefont{Sawicki}},
  \bibinfo{author}{\bibfnamefont{T.}~\bibnamefont{Dietl}},
  \bibnamefont{et~al.}, \bibinfo{journal}{Physical Review Letters}
  \textbf{\bibinfo{volume}{92}}, \bibinfo{pages}{037201}
  (\bibinfo{year}{2004}), ISSN \bibinfo{issn}{0031-9007},
  \urlprefix\url{http://link.aps.org/doi/10.1103/PhysRevLett.92.037201}.

\bibitem[{\citenamefont{Ku et~al.}(2003)\citenamefont{Ku, Potashnik, Wang,
  Chun, Schiffer, Samarth, Seong, Mascarenhas, Johnston-Halperin, Myers
  et~al.}}]{Ku2003}
\bibinfo{author}{\bibfnamefont{K.~C.} \bibnamefont{Ku}},
  \bibinfo{author}{\bibfnamefont{S.~J.} \bibnamefont{Potashnik}},
  \bibinfo{author}{\bibfnamefont{R.~F.} \bibnamefont{Wang}},
  \bibinfo{author}{\bibfnamefont{S.~H.} \bibnamefont{Chun}},
  \bibinfo{author}{\bibfnamefont{P.}~\bibnamefont{Schiffer}},
  \bibinfo{author}{\bibfnamefont{N.}~\bibnamefont{Samarth}},
  \bibinfo{author}{\bibfnamefont{M.~J.} \bibnamefont{Seong}},
  \bibinfo{author}{\bibfnamefont{a.}~\bibnamefont{Mascarenhas}},
  \bibinfo{author}{\bibfnamefont{E.}~\bibnamefont{Johnston-Halperin}},
  \bibinfo{author}{\bibfnamefont{R.~C.} \bibnamefont{Myers}},
  \bibnamefont{et~al.}, \bibinfo{journal}{Applied Physics Letters}
  \textbf{\bibinfo{volume}{82}}, \bibinfo{pages}{2302} (\bibinfo{year}{2003}),
  ISSN \bibinfo{issn}{00036951},
  \urlprefix\url{http://link.aip.org/link/APPLAB/v82/i14/p2302/s1\&Agg=doi}.

\bibitem[{\citenamefont{Boeck et~al.}(1996)\citenamefont{Boeck, Oesterholt, and
  Esch}}]{Boeck1996}
\bibinfo{author}{\bibfnamefont{J.~D.} \bibnamefont{Boeck}},
  \bibinfo{author}{\bibfnamefont{R.}~\bibnamefont{Oesterholt}},
  \bibnamefont{and} \bibinfo{author}{\bibfnamefont{A.~V.} \bibnamefont{Esch}},
  \bibinfo{journal}{Applied physics} \textbf{\bibinfo{volume}{68}},
  \bibinfo{pages}{2744} (\bibinfo{year}{1996}),
  \urlprefix\url{http://ieeexplore.ieee.org/xpls/abs\_all.jsp?arnumber=4888072%
}.

\bibitem[{\citenamefont{Myers et~al.}(2006)\citenamefont{Myers, Sheu, Jackson,
  Gossard, Schiffer, Samarth, and Awschalom}}]{Myers2006}
\bibinfo{author}{\bibfnamefont{R.~C.} \bibnamefont{Myers}},
  \bibinfo{author}{\bibfnamefont{B.~L.} \bibnamefont{Sheu}},
  \bibinfo{author}{\bibfnamefont{A.~W.} \bibnamefont{Jackson}},
  \bibinfo{author}{\bibfnamefont{A.~C.} \bibnamefont{Gossard}},
  \bibinfo{author}{\bibfnamefont{P.}~\bibnamefont{Schiffer}},
  \bibinfo{author}{\bibfnamefont{N.}~\bibnamefont{Samarth}}, \bibnamefont{and}
  \bibinfo{author}{\bibfnamefont{D.~D.} \bibnamefont{Awschalom}},
  \bibinfo{journal}{Phys. Rev. B} \textbf{\bibinfo{volume}{74}},
  \bibinfo{pages}{155203} (\bibinfo{year}{2006}), ISSN
  \bibinfo{issn}{1098-0121},
  \urlprefix\url{http://link.aps.org/doi/10.1103/PhysRevB.74.155203}.

\bibitem[{\citenamefont{Dressel and Gr\"{u}ner}(2002)}]{Dressel}
\bibinfo{author}{\bibfnamefont{M.}~\bibnamefont{Dressel}} \bibnamefont{and}
  \bibinfo{author}{\bibfnamefont{G.}~\bibnamefont{Gr\"{u}ner}},
  \emph{\bibinfo{title}{{Electrodynamics of solids}}}
  (\bibinfo{publisher}{Cambridge University Press},
  \bibinfo{address}{Cambridge, MA}, \bibinfo{year}{2002}).

\bibitem[{\citenamefont{Sinova et~al.}(2002)\citenamefont{Sinova, Jungwirth,
  Yang, Ku\v{c}era, and MacDonald}}]{Sinova2002}
\bibinfo{author}{\bibfnamefont{J.}~\bibnamefont{Sinova}},
  \bibinfo{author}{\bibfnamefont{T.}~\bibnamefont{Jungwirth}},
  \bibinfo{author}{\bibfnamefont{S.-R.} \bibnamefont{Yang}},
  \bibinfo{author}{\bibfnamefont{J.}~\bibnamefont{Ku\v{c}era}},
  \bibnamefont{and} \bibinfo{author}{\bibfnamefont{A.~H.}
  \bibnamefont{MacDonald}}, \bibinfo{journal}{Phys. Rev. B}
  \textbf{\bibinfo{volume}{66}}, \bibinfo{pages}{041202}
  (\bibinfo{year}{2002}), ISSN \bibinfo{issn}{0163-1829},
  \urlprefix\url{http://link.aps.org/doi/10.1103/PhysRevB.66.041202}.

\bibitem[{\citenamefont{Chapler et~al.}(2011)\citenamefont{Chapler, Myers,
  Mack, Frenzel, Pursley, Burch, Singley, Dattelbaum, Samarth, Awschalom
  et~al.}}]{Chapler2011}
\bibinfo{author}{\bibfnamefont{B.~C.} \bibnamefont{Chapler}},
  \bibinfo{author}{\bibfnamefont{R.~C.} \bibnamefont{Myers}},
  \bibinfo{author}{\bibfnamefont{S.}~\bibnamefont{Mack}},
  \bibinfo{author}{\bibfnamefont{A.}~\bibnamefont{Frenzel}},
  \bibinfo{author}{\bibfnamefont{B.~C.} \bibnamefont{Pursley}},
  \bibinfo{author}{\bibfnamefont{K.~S.} \bibnamefont{Burch}},
  \bibinfo{author}{\bibfnamefont{E.~J.} \bibnamefont{Singley}},
  \bibinfo{author}{\bibfnamefont{a.~M.} \bibnamefont{Dattelbaum}},
  \bibinfo{author}{\bibfnamefont{N.}~\bibnamefont{Samarth}},
  \bibinfo{author}{\bibfnamefont{D.~D.} \bibnamefont{Awschalom}},
  \bibnamefont{et~al.}, \bibinfo{journal}{Phys. Rev. B}
  \textbf{\bibinfo{volume}{84}}, \bibinfo{pages}{081203}
  (\bibinfo{year}{2011}), ISSN \bibinfo{issn}{1098-0121},
  \urlprefix\url{http://link.aps.org/doi/10.1103/PhysRevB.84.081203}.

\bibitem[{\citenamefont{Blinowski and Kacman}(2003)}]{Blinowski2003}
\bibinfo{author}{\bibfnamefont{J.}~\bibnamefont{Blinowski}} \bibnamefont{and}
  \bibinfo{author}{\bibfnamefont{P.}~\bibnamefont{Kacman}},
  \bibinfo{journal}{Physical Review B} \textbf{\bibinfo{volume}{67}},
  \bibinfo{pages}{120204(R)} (\bibinfo{year}{2003}), ISSN
  \bibinfo{issn}{0163-1829},
  \urlprefix\url{http://link.aps.org/doi/10.1103/PhysRevB.67.121204}.

\bibitem[{\citenamefont{Ma\v{s}ek and M\'{a}ca}(2004)}]{Masek2004}
\bibinfo{author}{\bibfnamefont{J.}~\bibnamefont{Ma\v{s}ek}} \bibnamefont{and}
  \bibinfo{author}{\bibfnamefont{F.}~\bibnamefont{M\'{a}ca}},
  \bibinfo{journal}{Physical Review B} \textbf{\bibinfo{volume}{69}},
  \bibinfo{pages}{165212} (\bibinfo{year}{2004}), ISSN
  \bibinfo{issn}{1098-0121},
  \urlprefix\url{http://link.aps.org/doi/10.1103/PhysRevB.69.165212}.

\bibitem[{\citenamefont{Bouzerar et~al.}(2005)\citenamefont{Bouzerar, Ziman,
  and Kudrnovsk\'{y}}}]{Bouzerar2005}
\bibinfo{author}{\bibfnamefont{G.}~\bibnamefont{Bouzerar}},
  \bibinfo{author}{\bibfnamefont{T.}~\bibnamefont{Ziman}}, \bibnamefont{and}
  \bibinfo{author}{\bibfnamefont{J.}~\bibnamefont{Kudrnovsk\'{y}}},
  \bibinfo{journal}{Physical Review B} \textbf{\bibinfo{volume}{72}},
  \bibinfo{pages}{125207} (\bibinfo{year}{2005}), ISSN
  \bibinfo{issn}{1098-0121},
  \urlprefix\url{http://link.aps.org/doi/10.1103/PhysRevB.72.125207}.

\bibitem[{\citenamefont{Takeda et~al.}(2008)\citenamefont{Takeda, Kobayashi,
  Okane, Ohkochi, Okamoto, Saitoh, Kobayashi, Yamagami, Fujimori, Tanaka
  et~al.}}]{Takeda2008}
\bibinfo{author}{\bibfnamefont{Y.}~\bibnamefont{Takeda}},
  \bibinfo{author}{\bibfnamefont{M.}~\bibnamefont{Kobayashi}},
  \bibinfo{author}{\bibfnamefont{T.}~\bibnamefont{Okane}},
  \bibinfo{author}{\bibfnamefont{T.}~\bibnamefont{Ohkochi}},
  \bibinfo{author}{\bibfnamefont{J.}~\bibnamefont{Okamoto}},
  \bibinfo{author}{\bibfnamefont{Y.}~\bibnamefont{Saitoh}},
  \bibinfo{author}{\bibfnamefont{K.}~\bibnamefont{Kobayashi}},
  \bibinfo{author}{\bibfnamefont{H.}~\bibnamefont{Yamagami}},
  \bibinfo{author}{\bibfnamefont{A.}~\bibnamefont{Fujimori}},
  \bibinfo{author}{\bibfnamefont{A.}~\bibnamefont{Tanaka}},
  \bibnamefont{et~al.}, \bibinfo{journal}{Physical Review Letters}
  \textbf{\bibinfo{volume}{100}}, \bibinfo{pages}{247202}
  (\bibinfo{year}{2008}), ISSN \bibinfo{issn}{0031-9007},
  \urlprefix\url{http://link.aps.org/doi/10.1103/PhysRevLett.100.247202}.

\bibitem[{\citenamefont{Zhu et~al.}(2007)\citenamefont{Zhu, Li, Xiang, and
  Samarth}}]{Zhu2007}
\bibinfo{author}{\bibfnamefont{M.}~\bibnamefont{Zhu}},
  \bibinfo{author}{\bibfnamefont{X.}~\bibnamefont{Li}},
  \bibinfo{author}{\bibfnamefont{G.}~\bibnamefont{Xiang}}, \bibnamefont{and}
  \bibinfo{author}{\bibfnamefont{N.}~\bibnamefont{Samarth}},
  \bibinfo{journal}{Physical Review B} \textbf{\bibinfo{volume}{76}},
  \bibinfo{pages}{201201} (\bibinfo{year}{2007}), ISSN
  \bibinfo{issn}{1098-0121},
  \urlprefix\url{http://link.aps.org/doi/10.1103/PhysRevB.76.201201}.

\bibitem[{\citenamefont{Chapler et~al.}(2012)\citenamefont{Chapler, Mack, Ju,
  Elson, Boudouris, Namdas, Yuen, Heeger, Samarth, {Di Ventra}
  et~al.}}]{Chapler2012}
\bibinfo{author}{\bibfnamefont{B.}~\bibnamefont{Chapler}},
  \bibinfo{author}{\bibfnamefont{S.}~\bibnamefont{Mack}},
  \bibinfo{author}{\bibfnamefont{L.}~\bibnamefont{Ju}},
  \bibinfo{author}{\bibfnamefont{T.}~\bibnamefont{Elson}},
  \bibinfo{author}{\bibfnamefont{B.}~\bibnamefont{Boudouris}},
  \bibinfo{author}{\bibfnamefont{E.}~\bibnamefont{Namdas}},
  \bibinfo{author}{\bibfnamefont{J.}~\bibnamefont{Yuen}},
  \bibinfo{author}{\bibfnamefont{A.}~\bibnamefont{Heeger}},
  \bibinfo{author}{\bibfnamefont{N.}~\bibnamefont{Samarth}},
  \bibinfo{author}{\bibfnamefont{M.}~\bibnamefont{{Di Ventra}}},
  \bibnamefont{et~al.}, \bibinfo{journal}{Physical Review B}
  \textbf{\bibinfo{volume}{86}}, \bibinfo{pages}{165302}
  (\bibinfo{year}{2012}), ISSN \bibinfo{issn}{1098-0121},
  \urlprefix\url{http://link.aps.org/doi/10.1103/PhysRevB.86.165302}.

\bibitem[{\citenamefont{Ohya et~al.}(2011{\natexlab{a}})\citenamefont{Ohya,
  Takata, and Tanaka}}]{Ohya2011}
\bibinfo{author}{\bibfnamefont{S.}~\bibnamefont{Ohya}},
  \bibinfo{author}{\bibfnamefont{K.}~\bibnamefont{Takata}}, \bibnamefont{and}
  \bibinfo{author}{\bibfnamefont{M.}~\bibnamefont{Tanaka}},
  \bibinfo{journal}{Nature Physics} \textbf{\bibinfo{volume}{7}},
  \bibinfo{pages}{342} (\bibinfo{year}{2011}{\natexlab{a}}), ISSN
  \bibinfo{issn}{1745-2473},
  \urlprefix\url{http://www.nature.com/doifinder/10.1038/nphys1905}.

\bibitem[{\citenamefont{Dobrowolska et~al.}(2012)\citenamefont{Dobrowolska,
  Tivakornsasithorn, Liu, Furdyna, Berciu, Yu, and
  Walukiewicz}}]{Dobrowolska2012}
\bibinfo{author}{\bibfnamefont{M.}~\bibnamefont{Dobrowolska}},
  \bibinfo{author}{\bibfnamefont{K.}~\bibnamefont{Tivakornsasithorn}},
  \bibinfo{author}{\bibfnamefont{X.}~\bibnamefont{Liu}},
  \bibinfo{author}{\bibfnamefont{J.~K.} \bibnamefont{Furdyna}},
  \bibinfo{author}{\bibfnamefont{M.}~\bibnamefont{Berciu}},
  \bibinfo{author}{\bibfnamefont{K.~M.} \bibnamefont{Yu}}, \bibnamefont{and}
  \bibinfo{author}{\bibfnamefont{W.}~\bibnamefont{Walukiewicz}},
  \bibinfo{journal}{Nature materials} \textbf{\bibinfo{volume}{11}},
  \bibinfo{pages}{444} (\bibinfo{year}{2012}), ISSN \bibinfo{issn}{1476-1122},
  \urlprefix\url{http://www.ncbi.nlm.nih.gov/pubmed/22344325}.

\bibitem[{\citenamefont{{Dobrowolska, M., Liu, X., Furdyna, J. K., Berciu, M.,
  Yu, K. M., Walukiewicz}}(2012)}]{Dobo2012}
\bibinfo{author}{\bibfnamefont{W.}~\bibnamefont{{Dobrowolska, M., Liu, X.,
  Furdyna, J. K., Berciu, M., Yu, K. M., Walukiewicz}}},
  \bibinfo{journal}{Arxiv preprint} pp. \bibinfo{pages}{1--8}
  (\bibinfo{year}{2012}), \eprint{1004.4446}.

\bibitem[{\citenamefont{{Edmonds, K. W., Gallagher, B.L., Wang, M., Rushforth,
  A. W. Makarovsky, O. Patane, A. Campion, P., Foxon. C. T., Novak, V.,
  Jungwirth}}(2012)}]{Edwards2012}
\bibinfo{author}{\bibfnamefont{T.}~\bibnamefont{{Edmonds, K. W., Gallagher,
  B.L., Wang, M., Rushforth, A. W. Makarovsky, O. Patane, A. Campion, P.,
  Foxon. C. T., Novak, V., Jungwirth}}}, \bibinfo{journal}{Arxiv preprint} pp.
  \bibinfo{pages}{1--6} (\bibinfo{year}{2012}), \eprint{1211.3860v1}.

\bibitem[{\citenamefont{Basov et~al.}(2011)\citenamefont{Basov, Averitt,
  van~der Marel, Dressel, and Haule}}]{Basov2011}
\bibinfo{author}{\bibfnamefont{D.}~\bibnamefont{Basov}},
  \bibinfo{author}{\bibfnamefont{R.}~\bibnamefont{Averitt}},
  \bibinfo{author}{\bibfnamefont{D.}~\bibnamefont{van~der Marel}},
  \bibinfo{author}{\bibfnamefont{M.}~\bibnamefont{Dressel}}, \bibnamefont{and}
  \bibinfo{author}{\bibfnamefont{K.}~\bibnamefont{Haule}},
  \bibinfo{journal}{Rev. Mod. Phys.} \textbf{\bibinfo{volume}{83}},
  \bibinfo{pages}{471} (\bibinfo{year}{2011}), ISSN \bibinfo{issn}{0034-6861},
  \urlprefix\url{http://link.aps.org/doi/10.1103/RevModPhys.83.471}.

\bibitem[{\citenamefont{Stone et~al.}(2008)\citenamefont{Stone, Alberi, Tardif,
  Beeman, Yu, Walukiewicz, and Dubon}}]{Stone2008}
\bibinfo{author}{\bibfnamefont{P.}~\bibnamefont{Stone}},
  \bibinfo{author}{\bibfnamefont{K.}~\bibnamefont{Alberi}},
  \bibinfo{author}{\bibfnamefont{S.}~\bibnamefont{Tardif}},
  \bibinfo{author}{\bibfnamefont{J.}~\bibnamefont{Beeman}},
  \bibinfo{author}{\bibfnamefont{K.}~\bibnamefont{Yu}},
  \bibinfo{author}{\bibfnamefont{W.}~\bibnamefont{Walukiewicz}},
  \bibnamefont{and} \bibinfo{author}{\bibfnamefont{O.}~\bibnamefont{Dubon}},
  \bibinfo{journal}{Physical Review Letters} \textbf{\bibinfo{volume}{101}},
  \bibinfo{pages}{087203} (\bibinfo{year}{2008}), ISSN
  \bibinfo{issn}{0031-9007},
  \urlprefix\url{http://link.aps.org/doi/10.1103/PhysRevLett.101.087203}.

\bibitem[{\citenamefont{Ando et~al.}(2008)\citenamefont{Ando, Saito, Agarwal,
  Debnath, and Zayets}}]{Ando2008}
\bibinfo{author}{\bibfnamefont{K.}~\bibnamefont{Ando}},
  \bibinfo{author}{\bibfnamefont{H.}~\bibnamefont{Saito}},
  \bibinfo{author}{\bibfnamefont{K.~C.} \bibnamefont{Agarwal}},
  \bibinfo{author}{\bibfnamefont{M.~C.} \bibnamefont{Debnath}},
  \bibnamefont{and} \bibinfo{author}{\bibfnamefont{V.}~\bibnamefont{Zayets}},
  \bibinfo{journal}{Phys. Rev. Lett.} \textbf{\bibinfo{volume}{100}},
  \bibinfo{pages}{067204} (\bibinfo{year}{2008}), ISSN
  \bibinfo{issn}{0031-9007},
  \urlprefix\url{http://link.aps.org/doi/10.1103/PhysRevLett.100.067204}.

\bibitem[{\citenamefont{Tang and Flatt\'{e}}(2008)}]{Tang2008}
\bibinfo{author}{\bibfnamefont{J.-M.} \bibnamefont{Tang}} \bibnamefont{and}
  \bibinfo{author}{\bibfnamefont{M.}~\bibnamefont{Flatt\'{e}}},
  \bibinfo{journal}{Phys. Rev. Lett.} \textbf{\bibinfo{volume}{101}},
  \bibinfo{pages}{157203} (\bibinfo{year}{2008}), ISSN
  \bibinfo{issn}{0031-9007},
  \urlprefix\url{http://link.aps.org/doi/10.1103/PhysRevLett.101.157203}.

\bibitem[{\citenamefont{Ma\v{s}ek et~al.}(2010)\citenamefont{Ma\v{s}ek,
  M\'{a}ca, Kudrnovsk\'{y}, Makarovsky, Eaves, Campion, Edmonds, Rushforth,
  Foxon, Gallagher et~al.}}]{Masek2010}
\bibinfo{author}{\bibfnamefont{J.}~\bibnamefont{Ma\v{s}ek}},
  \bibinfo{author}{\bibfnamefont{F.}~\bibnamefont{M\'{a}ca}},
  \bibinfo{author}{\bibfnamefont{J.}~\bibnamefont{Kudrnovsk\'{y}}},
  \bibinfo{author}{\bibfnamefont{O.}~\bibnamefont{Makarovsky}},
  \bibinfo{author}{\bibfnamefont{L.}~\bibnamefont{Eaves}},
  \bibinfo{author}{\bibfnamefont{R.}~\bibnamefont{Campion}},
  \bibinfo{author}{\bibfnamefont{K.}~\bibnamefont{Edmonds}},
  \bibinfo{author}{\bibfnamefont{A.}~\bibnamefont{Rushforth}},
  \bibinfo{author}{\bibfnamefont{C.}~\bibnamefont{Foxon}},
  \bibinfo{author}{\bibfnamefont{B.}~\bibnamefont{Gallagher}},
  \bibnamefont{et~al.}, \bibinfo{journal}{Physical Review Letters}
  \textbf{\bibinfo{volume}{105}}, \bibinfo{pages}{227202}
  (\bibinfo{year}{2010}), ISSN \bibinfo{issn}{0031-9007},
  \urlprefix\url{http://link.aps.org/doi/10.1103/PhysRevLett.105.227202}.

\bibitem[{\citenamefont{Sato and Dederics}(2003)}]{Sato2003}
\bibinfo{author}{\bibfnamefont{K.}~\bibnamefont{Sato}} \bibnamefont{and}
  \bibinfo{author}{\bibfnamefont{P.~H.} \bibnamefont{Dederics}},
  \bibinfo{journal}{Europhys. Lettl} \textbf{\bibinfo{volume}{61}},
  \bibinfo{pages}{403} (\bibinfo{year}{2003}),
  \urlprefix\url{http://iopscience.iop.org/0295-5075/61/3/403}.

\bibitem[{\citenamefont{Mack et~al.}(2008)\citenamefont{Mack, Myers, Heron,
  Gossard, and Awschalom}}]{Mack2008}
\bibinfo{author}{\bibfnamefont{S.}~\bibnamefont{Mack}},
  \bibinfo{author}{\bibfnamefont{R.~C.} \bibnamefont{Myers}},
  \bibinfo{author}{\bibfnamefont{J.~T.} \bibnamefont{Heron}},
  \bibinfo{author}{\bibfnamefont{A.~C.} \bibnamefont{Gossard}},
  \bibnamefont{and} \bibinfo{author}{\bibfnamefont{D.~D.}
  \bibnamefont{Awschalom}}, \bibinfo{journal}{Appl. Phys. Lett.}
  \textbf{\bibinfo{volume}{92}}, \bibinfo{pages}{192502}
  (\bibinfo{year}{2008}), ISSN \bibinfo{issn}{00036951},
  \urlprefix\url{http://link.aip.org/link/APPLAB/v92/i19/p192502/s1\&Agg=doi}.

\bibitem[{\citenamefont{Burch et~al.}(2004)\citenamefont{Burch, Stephens,
  Kawakami, Awschalom, and Basov}}]{Burch2004}
\bibinfo{author}{\bibfnamefont{K.~S.} \bibnamefont{Burch}},
  \bibinfo{author}{\bibfnamefont{J.}~\bibnamefont{Stephens}},
  \bibinfo{author}{\bibfnamefont{R.~K.} \bibnamefont{Kawakami}},
  \bibinfo{author}{\bibfnamefont{D.~D.} \bibnamefont{Awschalom}},
  \bibnamefont{and} \bibinfo{author}{\bibfnamefont{D.~N.} \bibnamefont{Basov}},
  \bibinfo{journal}{Phys. Rev. B} \textbf{\bibinfo{volume}{70}},
  \bibinfo{pages}{205208} (\bibinfo{year}{2004}), ISSN
  \bibinfo{issn}{1098-0121},
  \urlprefix\url{http://link.aps.org/doi/10.1103/PhysRevB.70.205208}.

\bibitem[{\citenamefont{Kuzmenko}(2005)}]{Kuzmenko2005}
\bibinfo{author}{\bibfnamefont{A.~B.} \bibnamefont{Kuzmenko}},
  \bibinfo{journal}{Review of Scientific Instruments}
  \textbf{\bibinfo{volume}{76}}, \bibinfo{pages}{083108}
  (\bibinfo{year}{2005}), ISSN \bibinfo{issn}{00346748},
  \urlprefix\url{http://link.aip.org/link/RSINAK/v76/i8/p083108/s1\&Agg=doi}.

\bibitem[{\citenamefont{Hirakawa et~al.}(2002)\citenamefont{Hirakawa,
  Katsumoto, Hayashi, Hashimoto, and Iye}}]{Hirakawa2002}
\bibinfo{author}{\bibfnamefont{K.}~\bibnamefont{Hirakawa}},
  \bibinfo{author}{\bibfnamefont{S.}~\bibnamefont{Katsumoto}},
  \bibinfo{author}{\bibfnamefont{T.}~\bibnamefont{Hayashi}},
  \bibinfo{author}{\bibfnamefont{Y.}~\bibnamefont{Hashimoto}},
  \bibnamefont{and} \bibinfo{author}{\bibfnamefont{Y.}~\bibnamefont{Iye}},
  \bibinfo{journal}{Physical Review B} \textbf{\bibinfo{volume}{65}},
  \bibinfo{pages}{4} (\bibinfo{year}{2002}), ISSN \bibinfo{issn}{0163-1829},
  \urlprefix\url{http://link.aps.org/doi/10.1103/PhysRevB.65.193312}.

\bibitem[{\citenamefont{Hirakawa}(2001)}]{Hirakawa2001}
\bibinfo{author}{\bibfnamefont{K.}~\bibnamefont{Hirakawa}},
  \bibinfo{journal}{Physica E: Low-dimensional Systems and Nanostructures}
  \textbf{\bibinfo{volume}{10}}, \bibinfo{pages}{215} (\bibinfo{year}{2001}),
  ISSN \bibinfo{issn}{13869477},
  \urlprefix\url{http://linkinghub.elsevier.com/retrieve/pii/S1386947701000856%
}.

\bibitem[{\citenamefont{Okimoto et~al.}(1995)\citenamefont{Okimoto, Katsufuji,
  Ishikawa, Urushibara, Arima, and Tokura}}]{Okimoto1995}
\bibinfo{author}{\bibfnamefont{Y.}~\bibnamefont{Okimoto}},
  \bibinfo{author}{\bibfnamefont{T.}~\bibnamefont{Katsufuji}},
  \bibinfo{author}{\bibfnamefont{T.}~\bibnamefont{Ishikawa}},
  \bibinfo{author}{\bibfnamefont{A.}~\bibnamefont{Urushibara}},
  \bibinfo{author}{\bibfnamefont{T.}~\bibnamefont{Arima}}, \bibnamefont{and}
  \bibinfo{author}{\bibfnamefont{Y.}~\bibnamefont{Tokura}},
  \bibinfo{journal}{Phys. Rev. Lett.} \textbf{\bibinfo{volume}{75}},
  \bibinfo{pages}{109} (\bibinfo{year}{1995}),
  \urlprefix\url{http://link.aps.org/doi/10.1103/PhysRevLett.75.109}.

\bibitem[{\citenamefont{Alberi et~al.}(2008)\citenamefont{Alberi, Yu, Stone,
  Dubon, Walukiewicz, Wojtowicz, Liu, and Furdyna}}]{Alberi2008}
\bibinfo{author}{\bibfnamefont{K.}~\bibnamefont{Alberi}},
  \bibinfo{author}{\bibfnamefont{K.}~\bibnamefont{Yu}},
  \bibinfo{author}{\bibfnamefont{P.}~\bibnamefont{Stone}},
  \bibinfo{author}{\bibfnamefont{O.}~\bibnamefont{Dubon}},
  \bibinfo{author}{\bibfnamefont{W.}~\bibnamefont{Walukiewicz}},
  \bibinfo{author}{\bibfnamefont{T.}~\bibnamefont{Wojtowicz}},
  \bibinfo{author}{\bibfnamefont{X.}~\bibnamefont{Liu}}, \bibnamefont{and}
  \bibinfo{author}{\bibfnamefont{J.}~\bibnamefont{Furdyna}},
  \bibinfo{journal}{Phys. Rev. B} \textbf{\bibinfo{volume}{78}},
  \bibinfo{pages}{075201} (\bibinfo{year}{2008}), ISSN
  \bibinfo{issn}{1098-0121},
  \urlprefix\url{http://link.aps.org/doi/10.1103/PhysRevB.78.075201}.

\bibitem[{\citenamefont{Wiley and {DiDomenico Jr.}}(1970)}]{Wiley1970}
\bibinfo{author}{\bibfnamefont{J.}~\bibnamefont{Wiley}} \bibnamefont{and}
  \bibinfo{author}{\bibfnamefont{M.}~\bibnamefont{{DiDomenico Jr.}}},
  \bibinfo{journal}{Physical Review B} \textbf{\bibinfo{volume}{2}},
  \bibinfo{pages}{427} (\bibinfo{year}{1970}),
  \urlprefix\url{http://prb.aps.org/abstract/PRB/v2/i2/p427\_1}.

\bibitem[{\citenamefont{Singley et~al.}(2003)\citenamefont{Singley, Burch,
  Kawakami, Stephens, Awschalom, and Basov}}]{Singley2003}
\bibinfo{author}{\bibfnamefont{E.~J.} \bibnamefont{Singley}},
  \bibinfo{author}{\bibfnamefont{K.~S.} \bibnamefont{Burch}},
  \bibinfo{author}{\bibfnamefont{R.~K.} \bibnamefont{Kawakami}},
  \bibinfo{author}{\bibfnamefont{J.}~\bibnamefont{Stephens}},
  \bibinfo{author}{\bibfnamefont{D.~D.} \bibnamefont{Awschalom}},
  \bibnamefont{and} \bibinfo{author}{\bibfnamefont{D.~N.} \bibnamefont{Basov}},
  \bibinfo{journal}{Phys. Rev. B} \textbf{\bibinfo{volume}{68}},
  \bibinfo{pages}{165204} (\bibinfo{year}{2003}), ISSN
  \bibinfo{issn}{0163-1829},
  \urlprefix\url{http://link.aps.org/doi/10.1103/PhysRevB.68.165204}.

\bibitem[{\citenamefont{Moca et~al.}(2009)\citenamefont{Moca, Zar\'{a}nd, and
  Berciu}}]{Moca2009a}
\bibinfo{author}{\bibfnamefont{C.~P.} \bibnamefont{Moca}},
  \bibinfo{author}{\bibfnamefont{G.}~\bibnamefont{Zar\'{a}nd}},
  \bibnamefont{and} \bibinfo{author}{\bibfnamefont{M.}~\bibnamefont{Berciu}},
  \bibinfo{journal}{Phys. Rev. B} \textbf{\bibinfo{volume}{80}},
  \bibinfo{pages}{165202} (\bibinfo{year}{2009}), ISSN
  \bibinfo{issn}{1098-0121},
  \urlprefix\url{http://link.aps.org/doi/10.1103/PhysRevB.80.165202}.

\bibitem[{\citenamefont{Bouzerar and Bouzerar}(2011)}]{Bouzerar2011a}
\bibinfo{author}{\bibfnamefont{G.}~\bibnamefont{Bouzerar}} \bibnamefont{and}
  \bibinfo{author}{\bibfnamefont{R.}~\bibnamefont{Bouzerar}},
  \bibinfo{journal}{New Journal of Physics} \textbf{\bibinfo{volume}{13}},
  \bibinfo{pages}{023002} (\bibinfo{year}{2011}), ISSN
  \bibinfo{issn}{1367-2630},
  \urlprefix\url{http://stacks.iop.org/1367-2630/13/i=2/a=023002?key=crossref.%
09b5e489bdaa7e56e9a3a11495d1bf37}.

\bibitem[{\citenamefont{Songprakob et~al.}(2002)\citenamefont{Songprakob,
  Zallen, Tsu, and Liu}}]{Songprakob2002}
\bibinfo{author}{\bibfnamefont{W.}~\bibnamefont{Songprakob}},
  \bibinfo{author}{\bibfnamefont{R.}~\bibnamefont{Zallen}},
  \bibinfo{author}{\bibfnamefont{D.~V.} \bibnamefont{Tsu}}, \bibnamefont{and}
  \bibinfo{author}{\bibfnamefont{W.~K.} \bibnamefont{Liu}},
  \bibinfo{journal}{J. Appl. Phys.} \textbf{\bibinfo{volume}{91}},
  \bibinfo{pages}{171} (\bibinfo{year}{2002}), ISSN \bibinfo{issn}{00218979},
  \urlprefix\url{http://link.aip.org/link/JAPIAU/v91/i1/p171/s1\&Agg=doi}.

\bibitem[{\citenamefont{Moriya and Munekata}(2003)}]{Moriya2003}
\bibinfo{author}{\bibfnamefont{R.}~\bibnamefont{Moriya}} \bibnamefont{and}
  \bibinfo{author}{\bibfnamefont{H.}~\bibnamefont{Munekata}},
  \bibinfo{journal}{J. Appl. Phys.} \textbf{\bibinfo{volume}{93}},
  \bibinfo{pages}{4603} (\bibinfo{year}{2003}), ISSN \bibinfo{issn}{00218979},
  \urlprefix\url{http://link.aip.org/link/JAPIAU/v93/i8/p4603/s1\&Agg=doi}.

\bibitem[{\citenamefont{Jungwirth et~al.}(2007)\citenamefont{Jungwirth, Sinova,
  MacDonald, Gallagher, Nov\'{a}k, Edmonds, Rushforth, Campion, Foxon, Eaves
  et~al.}}]{Jungwirth2007}
\bibinfo{author}{\bibfnamefont{T.}~\bibnamefont{Jungwirth}},
  \bibinfo{author}{\bibfnamefont{J.}~\bibnamefont{Sinova}},
  \bibinfo{author}{\bibfnamefont{A.~H.} \bibnamefont{MacDonald}},
  \bibinfo{author}{\bibfnamefont{B.~L.} \bibnamefont{Gallagher}},
  \bibinfo{author}{\bibfnamefont{V.}~\bibnamefont{Nov\'{a}k}},
  \bibinfo{author}{\bibfnamefont{K.~W.} \bibnamefont{Edmonds}},
  \bibinfo{author}{\bibfnamefont{A.~W.} \bibnamefont{Rushforth}},
  \bibinfo{author}{\bibfnamefont{R.~P.} \bibnamefont{Campion}},
  \bibinfo{author}{\bibfnamefont{C.~T.} \bibnamefont{Foxon}},
  \bibinfo{author}{\bibfnamefont{L.}~\bibnamefont{Eaves}},
  \bibnamefont{et~al.}, \bibinfo{journal}{Phys. Rev. B}
  \textbf{\bibinfo{volume}{76}}, \bibinfo{pages}{125206}
  (\bibinfo{year}{2007}), ISSN \bibinfo{issn}{1098-0121},
  \urlprefix\url{http://link.aps.org/doi/10.1103/PhysRevB.76.125206}.

\bibitem[{\citenamefont{Bebb}(1967)}]{Bebb1967}
\bibinfo{author}{\bibfnamefont{H.}~\bibnamefont{Bebb}},
  \bibinfo{journal}{Journal of Physics and Chemistry of Solids}
  \textbf{\bibinfo{volume}{28}}, \bibinfo{pages}{2087} (\bibinfo{year}{1967}),
  \urlprefix\url{http://linkinghub.elsevier.com/retrieve/pii/0022369767901849}.

\bibitem[{\citenamefont{Bebb}(1969)}]{Bebb1969}
\bibinfo{author}{\bibfnamefont{H.}~\bibnamefont{Bebb}},
  \bibinfo{journal}{Physical Review} \textbf{\bibinfo{volume}{185}},
  \bibinfo{pages}{1116} (\bibinfo{year}{1969}), ISSN \bibinfo{issn}{0031-899X},
  \urlprefix\url{http://link.aps.org/doi/10.1103/PhysRev.185.1116}.

\bibitem[{\citenamefont{Huber et~al.}(1987)\citenamefont{Huber, Perez, and
  Huber}}]{Huber1987}
\bibinfo{author}{\bibfnamefont{C.}~\bibnamefont{Huber}},
  \bibinfo{author}{\bibfnamefont{J.}~\bibnamefont{Perez}}, \bibnamefont{and}
  \bibinfo{author}{\bibfnamefont{T.}~\bibnamefont{Huber}},
  \bibinfo{journal}{Physical Review B} \textbf{\bibinfo{volume}{36}},
  \bibinfo{pages}{5933} (\bibinfo{year}{1987}), ISSN \bibinfo{issn}{1550-235X},
  \urlprefix\url{http://link.aps.org/doi/10.1103/PhysRevB.36.5933}.

\bibitem[{\citenamefont{Kojima et~al.}(2007)\citenamefont{Kojima, H\'{e}roux,
  Shimano, Hashimoto, Katsumoto, Iye, and Kuwata-Gonokami}}]{Kojima2007}
\bibinfo{author}{\bibfnamefont{E.}~\bibnamefont{Kojima}},
  \bibinfo{author}{\bibfnamefont{J.}~\bibnamefont{H\'{e}roux}},
  \bibinfo{author}{\bibfnamefont{R.}~\bibnamefont{Shimano}},
  \bibinfo{author}{\bibfnamefont{Y.}~\bibnamefont{Hashimoto}},
  \bibinfo{author}{\bibfnamefont{S.}~\bibnamefont{Katsumoto}},
  \bibinfo{author}{\bibfnamefont{Y.}~\bibnamefont{Iye}}, \bibnamefont{and}
  \bibinfo{author}{\bibfnamefont{M.}~\bibnamefont{Kuwata-Gonokami}},
  \bibinfo{journal}{Phys. Rev. B} \textbf{\bibinfo{volume}{76}},
  \bibinfo{pages}{195323} (\bibinfo{year}{2007}), ISSN
  \bibinfo{issn}{1098-0121},
  \urlprefix\url{http://link.aps.org/doi/10.1103/PhysRevB.76.195323}.

\bibitem[{\citenamefont{Nagai and Nagasaka}(2006)}]{Nagai2005}
\bibinfo{author}{\bibfnamefont{Y.}~\bibnamefont{Nagai}} \bibnamefont{and}
  \bibinfo{author}{\bibfnamefont{K.}~\bibnamefont{Nagasaka}},
  \bibinfo{journal}{Infrared Phys. Techno} \textbf{\bibinfo{volume}{48}},
  \bibinfo{pages}{1} (\bibinfo{year}{2006}).

\bibitem[{\citenamefont{Braunstein and Kane}(1962)}]{Braunstein1962}
\bibinfo{author}{\bibfnamefont{R.}~\bibnamefont{Braunstein}} \bibnamefont{and}
  \bibinfo{author}{\bibfnamefont{E.~O.} \bibnamefont{Kane}},
  \bibinfo{journal}{J. Phys. Chem. Solids} \textbf{\bibinfo{volume}{23}},
  \bibinfo{pages}{1423} (\bibinfo{year}{1962}),
  \urlprefix\url{http://linkinghub.elsevier.com/retrieve/pii/0022369762901956}.

\bibitem[{\citenamefont{Okabayashi et~al.}(2001)\citenamefont{Okabayashi,
  Kimura, Rader, Mizokawa, Fujimori, Hayashi, and Tanaka}}]{Okabayashi2001}
\bibinfo{author}{\bibfnamefont{J.}~\bibnamefont{Okabayashi}},
  \bibinfo{author}{\bibfnamefont{A.}~\bibnamefont{Kimura}},
  \bibinfo{author}{\bibfnamefont{O.}~\bibnamefont{Rader}},
  \bibinfo{author}{\bibfnamefont{T.}~\bibnamefont{Mizokawa}},
  \bibinfo{author}{\bibfnamefont{A.}~\bibnamefont{Fujimori}},
  \bibinfo{author}{\bibfnamefont{T.}~\bibnamefont{Hayashi}}, \bibnamefont{and}
  \bibinfo{author}{\bibfnamefont{M.}~\bibnamefont{Tanaka}},
  \bibinfo{journal}{Phys. Rev. B} \textbf{\bibinfo{volume}{64}},
  \bibinfo{pages}{125304} (\bibinfo{year}{2001}), ISSN
  \bibinfo{issn}{0163-1829},
  \urlprefix\url{http://link.aps.org/doi/10.1103/PhysRevB.64.125304}.

\bibitem[{\citenamefont{Mayer et~al.}(2010)\citenamefont{Mayer, Stone, Miller,
  Smith, Dubon, Haller, Yu, Walukiewicz, Liu, and Furdyna}}]{Mayer2010}
\bibinfo{author}{\bibfnamefont{M.~A.} \bibnamefont{Mayer}},
  \bibinfo{author}{\bibfnamefont{P.~R.} \bibnamefont{Stone}},
  \bibinfo{author}{\bibfnamefont{N.}~\bibnamefont{Miller}},
  \bibinfo{author}{\bibfnamefont{H.~M.} \bibnamefont{Smith}},
  \bibinfo{author}{\bibfnamefont{O.~D.} \bibnamefont{Dubon}},
  \bibinfo{author}{\bibfnamefont{E.~E.} \bibnamefont{Haller}},
  \bibinfo{author}{\bibfnamefont{K.~M.} \bibnamefont{Yu}},
  \bibinfo{author}{\bibfnamefont{W.}~\bibnamefont{Walukiewicz}},
  \bibinfo{author}{\bibfnamefont{X.}~\bibnamefont{Liu}}, \bibnamefont{and}
  \bibinfo{author}{\bibfnamefont{J.~K.} \bibnamefont{Furdyna}},
  \bibinfo{journal}{Phys. Rev. B} \textbf{\bibinfo{volume}{81}},
  \bibinfo{pages}{045205} (\bibinfo{year}{2010}), ISSN
  \bibinfo{issn}{1098-0121},
  \urlprefix\url{http://link.aps.org/doi/10.1103/PhysRevB.81.045205}.

\bibitem[{\citenamefont{Ohya et~al.}(2010)\citenamefont{Ohya, Muneta, Hai, and
  Tanaka}}]{Ohya2010a}
\bibinfo{author}{\bibfnamefont{S.}~\bibnamefont{Ohya}},
  \bibinfo{author}{\bibfnamefont{I.}~\bibnamefont{Muneta}},
  \bibinfo{author}{\bibfnamefont{P.~N.} \bibnamefont{Hai}}, \bibnamefont{and}
  \bibinfo{author}{\bibfnamefont{M.}~\bibnamefont{Tanaka}},
  \bibinfo{journal}{Phys. Rev. Lett.} \textbf{\bibinfo{volume}{104}},
  \bibinfo{pages}{167204} (\bibinfo{year}{2010}), ISSN
  \bibinfo{issn}{0031-9007},
  \urlprefix\url{http://link.aps.org/doi/10.1103/PhysRevLett.104.167204}.

\bibitem[{\citenamefont{Dietl and Sztenkiel}(2011)}]{Dietl2011}
\bibinfo{author}{\bibfnamefont{T.}~\bibnamefont{Dietl}} \bibnamefont{and}
  \bibinfo{author}{\bibfnamefont{D.}~\bibnamefont{Sztenkiel}},
  \bibinfo{journal}{Arxiv preprint}  (\bibinfo{year}{2011}),
  \eprint{1102.3267v2}.

\bibitem[{\citenamefont{Ohya et~al.}(2011{\natexlab{b}})\citenamefont{Ohya,
  Takata, Muneta, Hai, and Tanaka}}]{Ohya2011B}
\bibinfo{author}{\bibfnamefont{S.}~\bibnamefont{Ohya}},
  \bibinfo{author}{\bibfnamefont{K.}~\bibnamefont{Takata}},
  \bibinfo{author}{\bibfnamefont{I.}~\bibnamefont{Muneta}},
  \bibinfo{author}{\bibfnamefont{P.~N.} \bibnamefont{Hai}}, \bibnamefont{and}
  \bibinfo{author}{\bibfnamefont{M.}~\bibnamefont{Tanaka}},
  \bibinfo{journal}{Arxiv preprint}  (\bibinfo{year}{2011}{\natexlab{b}}),
  \eprint{1102.4459}.

\bibitem[{\citenamefont{Sawicki et~al.}(2009)\citenamefont{Sawicki, Chiba,
  Korbecka, Nishitani, Majewski, Matsukura, Dietl, and Ohno}}]{Sawicki2009}
\bibinfo{author}{\bibfnamefont{M.}~\bibnamefont{Sawicki}},
  \bibinfo{author}{\bibfnamefont{D.}~\bibnamefont{Chiba}},
  \bibinfo{author}{\bibfnamefont{A.}~\bibnamefont{Korbecka}},
  \bibinfo{author}{\bibfnamefont{Y.}~\bibnamefont{Nishitani}},
  \bibinfo{author}{\bibfnamefont{J.~a.} \bibnamefont{Majewski}},
  \bibinfo{author}{\bibfnamefont{F.}~\bibnamefont{Matsukura}},
  \bibinfo{author}{\bibfnamefont{T.}~\bibnamefont{Dietl}}, \bibnamefont{and}
  \bibinfo{author}{\bibfnamefont{H.}~\bibnamefont{Ohno}},
  \bibinfo{journal}{Nat. Phys.} \textbf{\bibinfo{volume}{6}},
  \bibinfo{pages}{22} (\bibinfo{year}{2009}), ISSN \bibinfo{issn}{1745-2473},
  \urlprefix\url{http://www.nature.com/doifinder/10.1038/nphys1455}.

\bibitem[{\citenamefont{Richardella et~al.}(2010)\citenamefont{Richardella,
  Roushan, Mack, Zhou, Huse, Awschalom, and Yazdani}}]{Richardella2010}
\bibinfo{author}{\bibfnamefont{A.}~\bibnamefont{Richardella}},
  \bibinfo{author}{\bibfnamefont{P.}~\bibnamefont{Roushan}},
  \bibinfo{author}{\bibfnamefont{S.}~\bibnamefont{Mack}},
  \bibinfo{author}{\bibfnamefont{B.}~\bibnamefont{Zhou}},
  \bibinfo{author}{\bibfnamefont{D.~a.} \bibnamefont{Huse}},
  \bibinfo{author}{\bibfnamefont{D.~D.} \bibnamefont{Awschalom}},
  \bibnamefont{and} \bibinfo{author}{\bibfnamefont{A.}~\bibnamefont{Yazdani}},
  \bibinfo{journal}{Science (New York, N.Y.)} \textbf{\bibinfo{volume}{327}},
  \bibinfo{pages}{665} (\bibinfo{year}{2010}), ISSN \bibinfo{issn}{1095-9203},
  \urlprefix\url{http://www.ncbi.nlm.nih.gov/pubmed/20133566}.

\end{thebibliography}
\end{document}